\documentclass[journal,comsoc,twocolumn]{IEEEtran}

\usepackage[cmintegrals]{newtxmath}
\usepackage{epsfig}
\usepackage{afterpage}
\usepackage{graphicx}
\usepackage{amsmath}
\usepackage{cases}
\usepackage[caption=false,labelformat=simple]{subfig}
\usepackage{algorithm,algorithmic}
\usepackage[capitalise]{cleveref}

\usepackage{amsfonts}
\usepackage{cite}
\usepackage{times}
\usepackage{color}
\usepackage{cite}
\usepackage{psfrag}
\usepackage{multirow}
\usepackage{cuted,flushend}
\usepackage{mathtools}
\usepackage{float}
\usepackage{array}
\usepackage{hhline}
\usepackage{stfloats}
\usepackage{bm}
\usepackage[bottom]{footmisc}
\usepackage{mwe}
\usepackage{color}
\usepackage{booktabs,multirow}
\usepackage{tabularx}
\usepackage{tikz}
\usetikzlibrary{patterns,arrows}

\definecolor{color1}{RGB}{228,26,28}
\definecolor{color2}{RGB}{55,126,184}
\definecolor{color3}{RGB}{77,175,74}
\definecolor{color4}{RGB}{152,78,163}
\definecolor{color5}{RGB}{255,127,0}

\usepackage{nopageno}

\begin{document}
\bstctlcite{IEEEexample:BSTcontrol}

\title{\fontsize{22}{28}\selectfont Joint Association and Resource Allocation for Multi-Hop Integrated Access and Backhaul (IAB) Network}

\author{Byungju Lim, Ju-Hyung Lee, Jae-Hong Kwon, and~Young-Chai~Ko

\thanks{B. Lim, J. -H. Lee, J. -H. Kwon, and Y. -C. Ko are with the School of Electrical Engineering, Korea University, Seoul 02841, South Korea (E-mail: $<$limbj93, leejuhyung, hugokwon, koyc$>$@korea.ac.kr)}
\thanks{Part of this work was submitted to IEEE ICTC, Oct, 2021.}
}

\maketitle
\thispagestyle{empty}
\begin{abstract}
Integrated access and backhaul (IAB) network is envisioned as a novel network architecture for increasing the network capacity and coverage. To facilitate the IAB network, the appropriate methods of wireless link association and resource management are required.
In this paper, we investigate the joint optimization problem of association and resource allocation in terms of subchannel and power for IAB network. In particular, we handle the association and resource allocation problems for wireless backhaul and access links considering multi-hop backhauling.
Since the optimization problem for IAB network is formulated as a mixed integer non-linear programming (MINLP), we divide it into three subproblems for association, subchannel allocation, and power allocation, respectively, and these subproblems are solved alternatively to obtain a local optimal solution.
For the association problem, we adopt the Lagrangian duality approach to configure the backhaul and access links and successive convex approximation (SCA) approach is used to solve the subchannel and power allocation problems efficiently.
Simulation results demonstrate that the proposed algorithm achieves better performance than single-hop backhauling based network and enhances the capacity and coverage by configuring the multi-hop backhauling. 

\end{abstract}


\IEEEpeerreviewmaketitle

\section{Introduction}\label{sec:Intro}

The fifth generation (5G) and beyond 5G networks are envisioned to increase the capacity by 1000 fold for the rapid growth of data traffic.
To cope with exponentially increasing traffic demands, massive multiple-input multiple-output (MIMO) and mmWave technologies are considered as major capacity enhancing techniques, which exploits the benefits of spatial reuse and wide bandwidth \cite{what_5G}.
However, to take the advantage of mmWave communications, we need to overcome several challenges such as high propagation loss and high sensitivity to blockages, which makes the network coverage limited \cite{mmwave}. 
For seamless coverage in mmWave communications, network densification where the low-power small cell base stations (SBSs) are densely deployed, has been regarded as a possible approach \cite{mmwave_sbs}.


Heterogeneous network (HetNet) where SBSs are overlaid within macro BS (MBS) requires a backhaul connection between the core networks and SBSs \cite{HetNet}.
Tradition approach for backhaul connection is wired backhauling which provides high speed and reliable communications.
However, implementing the optical fiber backhaul link for the large-scale deployment of SBSs is infeasible due to the prohibitive cost.
In this respect, mmWave based wireless backhauling has been considered as an alternative that can be scalable in the network densification and provide the gigabit per second (Gbps) data rates \cite{mmwave_BH}.
Wireless backhauling effectively supports ever-increasing backhaul traffic demands while it has benefits in both hardware costs and deployment difficulties. 
Furthermore, using the beamforming technique, spatial reuse can be exploited in the wireless backhauling such that it can provide the immunity against inter-cell interference \cite{mmwave_BH}.
However, the extra radio resources and infrastructure are essential for implementing the mmWave based wireless backhaul.

In-band wireless backhaul, referred to as self-backhauling, utilizes the same resources and infrastructure for the access link \cite{In_band}. 
Accordingly, it does not require the extra spectrum for backhaul link and the existing infrastructure can be used to serve backhaul as well as access link.
With the advantages of in-band wireless backhaul, 3GPP introduces the integrated access and backhaul (IAB) which has been standardized in 5G Release 16 \cite{IAB_standard}.
According to \cite{IAB_standard}, the main characteristics of IAB are the integration of wireless access and backhaul links, the use of mmWave spectrum, and the plug-and-play installation of IAB nodes.

Multi-hop (MH) backhauling, one of the advantages of IAB technology, is a promising solution to enhance the network throughput and coverage. By exploiting the spatial reuse in IAB with MH backhauling, the blockage issue and the interference management in IAB network need to be efficiently addressed. Hence, the design of MH backhauling including topology management, route selection, and dynamic resource allocation between backhaul and access links, is the important and challenging issue in the IAB network \cite{IAB_magazine}.
However, designing the MH-IAB network may be difficult due to the flexibility of deployment and configuration.
Thus, efficient network design in-between access and backhaul links for MH-IAB is demanded while considering the interference caused by in-band wireless backhaul and blockage issues.

\subsection{Related work}

Several works investigated the performance of IAB network based on stochastic geometry \cite{BW_part,mm_IAB}. 
In particular, authors in \cite{BW_part,mm_IAB} analyzed the effects of bandwidth partition between access and backhaul links and presented an optimal bandwidth partition ratio for access and backhaul links.
In \cite{RA_markov,RA_hetnet,full_RA,joint_UA_RA}, the resource allocation for HetNet has been investigated. 
The resource management for time and frequency has been designed for maximizing the sum throughput under the existence of cross-tier interference in \cite{RA_markov,RA_hetnet}. 
Authors in \cite{full_RA} investigated the spectrum allocation problem between access and backhaul links for in-band and out-band full-duplex based SBSs.
Besides in \cite{joint_UA_RA}, the joint association and bandwidth allocation problem for wireless backhaul studied for two-tier HetNets.
However, these works only consider single-hop (SH) backhauling where all the SBSs are directly connected to MBS, such that the advantages of IAB network are not fully exploited.

MH backhauling has been recently investigated in \cite{traffic,inter_RA,Path_sel} to guarantee the coverage and the line-of-sight (LoS) condition for backhaul.
In \cite{traffic}, the resource allocation for backhaul was studied to configure the MH backhauling under fixed backhaul traffic.
Besides in \cite{inter_RA}, the framework of MH network for the resource management was addressed based on matching game theory. 
Moreover, optimal strategies for backhaul link selection were investigated in \cite{Path_sel} to configure MH backhauling where high quality first (HQF) selects the link with the highest SNR as a backhaul and wired first (WF) selects a direct link to MBS with the highest SNR above a threshold.
Despite of the studies for MH backhauling, all of these works do not consider the user traffics which significantly affect the overall network performance and the optimal backhaul link configuration.
Thus, it cannot provide the adaptive backhaul configuration under different traffic requirements of users.



\subsection{Main Contribution}
As aforementioned, the MH backhauling is a promising solution for the coverage and the blockage issues in IAB network. 
It is vital to efficiently design the MH backhauling and allocate resources between access and backhaul links.
To the best of our knowledge, the joint design of resource allocation and association for the MH backhauling while considering the traffics of users has not been investigated.
In this paper, for MH-IAB network, we jointly optimize the association and resource allocation with respect to subchannel and power taking into account of the dynamic user traffic demands.
Our proposed algorithm presents the optimized association of backhaul and access links and improves the network throughput by allocating the subchannel and power for each link. 
Our contributions are summarized as follows:
\begin{enumerate}

\item We propose the joint resource allocation and association optimization to enhance the network throughput while guaranteeing the quality of service (QoS) of users. 
In particular, we consider inter-cell interference and cross-tier interference caused by shared resources between SBSs and MBS.
     
\item  To solve the optimization problem, we propose the three stages based iterative algorithm. 
Firstly, the MH backhauling based optimal association is proposed using the Lagrangian duality approach.
We observe that the blockage issue, which is a major obstacle for the coverage, is circumvented by configuring the MH backhauling and it achieves higher LoS probability for backhaul links.
      
\item In the second and third stages, the subchannel and power allocation algorithms are proposed, respectively. 
We transform them into a sequence of convex problems based on SCA, which are the lower bound of the original problem, and the low complexity algorithms are proposed to solve these subproblems. 
Particularly, the mixed integer programming of subchannel allocation is addressed by employing the continuous relaxation and penalty function.
As a result, some simulation results confirm that the proposed algorithm provides better throughput and coverage performance than SH backhauling.

\end{enumerate}
   

The rest of this paper is organized as follows. Section~\ref{system} presents the system and channel model for IAB network and the joint optimization problem is formulated in Section \ref{problem}. In Section \ref{solution}, we solve the joint optimization problem by dividing it into three subproblems and the complexity of proposed algorithm is analyzed. 
The simulation results are discussed in Section \ref{simul} to verify the performance of proposed scheme and finally conclusion is provided in Section \ref{conclu}.

\section{System Model}\label{system}

\begin{figure}[t]
    \centering
    \includegraphics[width=85mm,height=50mm]{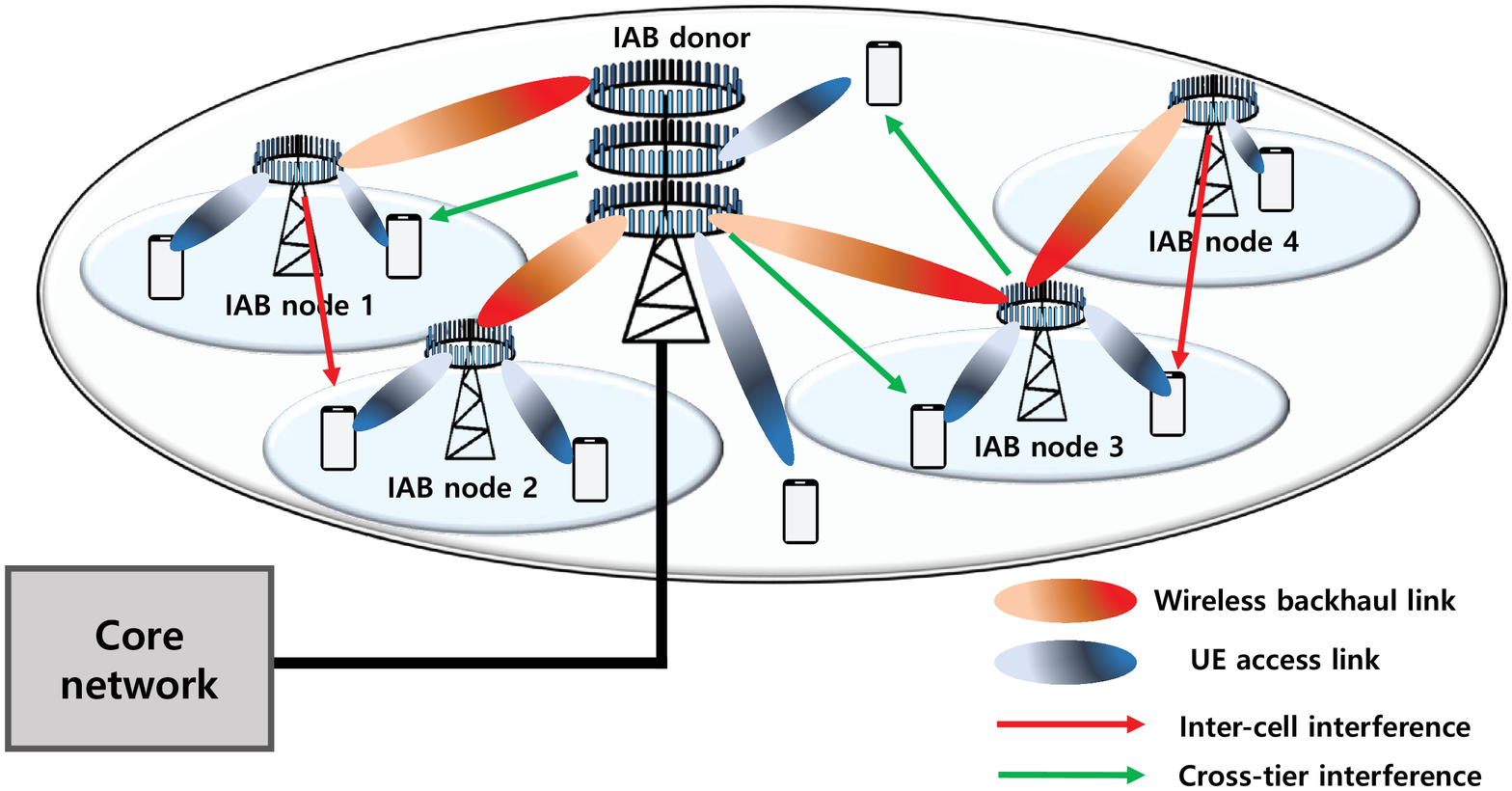}
    \caption{Multi-hop IAB based two-tier heterogeneous network.}\label{fig:system}
\end{figure}

\begin{figure}[t]
    \centering
    \includegraphics[width=85mm]{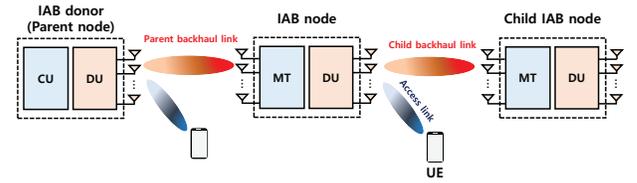}
    \caption{Basic architecture of multi-hop IAB network.}\label{fig:system2}
\end{figure}

\subsection{IAB Network Scenario}

Consider a two-tier HetNet where $B$ SBSs are distributed in a macrocell with MBS at its center.
We assume that MBS and SBS are equipped with $N_{m}$ and $N_{s}$ antenna arrays, respectively, and 
$K$ user equipments (UEs) equipped with $N_u$ antenna arrays are served by MBS or SBS.
It is also assumed that MBS is connected to core networks with high speed optical fiber and provides wireless backhaul connectivity for SBSs that are connected to the core networks through MBS. Note that wireless backhaul can be supported between SBSs since IAB can provide MH wireless backhauling as illustrated in  
\cref{fig:system} \cite{IAB_standard}.
Throughout this paper, IAB-donor is referred to as MBS and SBS indicates an IAB node since SBS acts as an IAB node.

IAB node consists of a distributed unit (DU) and a mobile terminal (MT) to simultaneously support both backhaul and access links. 
In \cref{fig:system2}, DU is used to serve the child IAB node and its associated UE, while MT is used to maintain the wireless backhaul connection to the parent node \cite{IAB_magazine}. 
Thus, we refer to the link between UE and IAB-DU as access link, the link between the IAB-MT and its parent node as parent backhaul link, and the link between the IAB-DU and its child IAB-MT as child backhaul link. 
For the SH backhauling, IAB-donor always acts as the parent node of all the IAB nodes. 
On the other hand, in MH backhauling, not only IAB-donor but also IAB node can serve as the parent node.
It should be noted that both of backhaul and access links are supported by the same wireless resources and all the backhaul and access links use mmWave spectrum.

For the mathematical convenience, $\mathcal{B}_0=\{0\} \bigcup \mathcal{B}$ denotes the set of all BSs, in which the index $0$ represents MBS while $\mathcal{B}=\{1,2,...,B\}$ represents the set of SBSs (e.g., IAB nodes).
Let $\mathcal{I}=\mathcal{B} \bigcup \mathcal{K}$ denote the set of all SBSs and UEs where $\mathcal{K}=\{B+1,B+2,...,B+K\}$ is the set of UEs. 
The set of subchannels denoted as $\mathcal{M=}\{1,2,...,M\}$ comprises with $M$ subchannels. 
The notation used throughout this paper is summarized in \cref{tab_notation}.

\begin{table}[t]
\centering
\caption{Notation of variables}\label{tab_notation}
\begin{tabular}{|m{0.12\textwidth}||m{0.31\textwidth}|}
\hline
\textbf{Notation} & \textbf{Definition}  \\ 
\hhline{|=#=|}
$B$ & Number of SBSs\\
\hline
$K$ & Number of Users\\
\hline
$M$ & Number of subchannels\\
\hline
$N_m$ & Number of antennas for MBS\\
\hline
$N_s$ & Number of antennas for SBS\\
\hline
$N_u$ & Number of antennas for UE\\
\hline
$L$ & Number of NLoS paths\\
\hline
$\mathcal{B}$, $\mathcal{B}_0$  & Sets of SBSs and all BSs including MBS\\
\hline
$\mathcal{K}$ & Set of UEs\\\hline
$\mathcal{I}$ & Set of SBSs and UEs\\\hline
$\mathcal{M}$ & Set of subchannels\\\hline
$R_i^\textrm{th}$ & Minimum data rate requirement of the $i$-th UE\\\hline
$P_b^\textrm{max}$ & Maximum transmit power at the $b$-th BS\\\hline
    $\beta_\textrm{LoS}$, $\beta_\textrm{NLoS}$ & Path loss exponent for LoS and NLOS paths\\
\hline
$\sigma_\textrm{LoS}$,$\sigma_\textrm{NLoS}$ & Standard deviation of shadow fading for LoS and NLoS paths\\
\hline
$\gamma_{b,i,m}$ & SINR of the $i$-th node from the $b$-th BS over the $m$-th subchannel\\
\hline
$\mathbf{H}_{b,i,m}$ & Channel matrix for the $i$-th node from the $b$-th BS over the $m$-th subchannel\\\hline
$\mathbf{v}_{b,i,m}$, $\mathbf{w}_{b,i,m}$ & Precoding and combining vector for the $i$-th node from the $b$-th BS over the $m$-th subchannel.\\\hline

$x_{b,i,m}$ & Subchannel allocation variable for the $i$-th node from the $b$-th BS over the $m$-th subchannel \\ 
\hline
$y_{b,i}$ & Association variable for the $i$-th node from the $b$-th BS over the $m$-th subchannel \\  
\hline
$P_{b,m}$ & Power allocation variable for the $b$-th BS over the $m$-th subchannel\\
\hline
\end{tabular}

\end{table}

\subsection{Channel Model}
\subsubsection{LoS and NLoS Channel models}
We consider a well-known frequency-dependent path loss (PL) model, namely the close-in free space reference distance PL (CI-PL) model \cite{PL_model}, which is given by 
\begin{align}
    \textrm{PL}(\beta,d)~ \textrm{[dB]}=a+10\beta\log_{10}d_{\textrm{[m]}}+\chi,
\end{align}
where $a=20\log_{10}\left(4\pi f_{c}/c\right)$ is the free space path loss with 1m reference distance, $\beta$ is the path loss exponent, and $\chi$ is the log-normal shadow fading. 
Note that $f_{c}$ and $c$ are the carrier frequency and the speed of light, respectively. 

Since the path loss exponent depends on whether there is a LoS path or not, the CI-PL models for LoS and NLoS are separately defined as $\textrm{PL}_{\textrm{LoS}}\left(d\right)=\textrm{PL}\left(\beta_\textrm{LoS},d\right)$ and $\textrm{PL}_{\textrm{NLoS}}\left(d\right)=\textrm{PL}\left(\beta_\textrm{NLoS},d\right)$, respectively.
Furthermore, the small-scale fading channels for LoS and NLoS paths are given respectively as
\begin{align}\notag
    \mathbf{H}^{\textrm{LoS}}_{b,i,m}&=\sqrt{N_{t}N_{r}}\alpha^{\textrm{LoS}}_{b,i,m}\mathbf{a}\left(\theta_{b,i}\right)\mathbf{a}^{H}\left(\phi_{b,i}\right),\\
    \mathbf{H}^{\textrm{NLoS}}_{b,i,m}&=\sqrt{\frac{N_{t}N_{r}}{L}}\sum_{l=1}^{L}\alpha^{\textrm{NLoS}}_{b,i,m,l}\mathbf{a}\left(\theta_{b,i,l}\right)\mathbf{a}^{H}\left(\phi_{b,i,l}\right),
\end{align}
where $\alpha^{\textrm{LoS}}_{b,i},\alpha^{\textrm{NLoS}}_{b,i}\sim CN\left(0,1\right)$, are the channel gain for LoS and NLoS, respectively, $\theta_{b,i}$ and $\phi_{b,i}$ are the angle of departure (AoD) and angle of arrival (AoA) from the $b$-th BS to the $i$-th node, respectively, $L$ is the number of NLoS paths, and $N_t$ and $N_r$ denote the number of Tx and Rx antenna arrays, respectively.
Since both MBS and SBS are equipped with uniform linear array (ULA) configuration, the array response vector of the ULA is defined as
\begin{align}
    \mathbf{a}\left(\theta\right)=\frac{1}{\sqrt{N}}\left[1,e^{j\frac{2\pi}{\lambda}d_a\sin\theta},\dots,e^{j\frac{2\pi}{\lambda}\left(N-1\right)d_a\sin\theta}\right]^{T},
\end{align}
where $N$, $\theta$, $\lambda$ and $d_a$ are the size of antenna array, the angle of arrival (or departure), the wavelength, and the antenna spacing, respectively. 
Thus, the channel matrix of the $b$-th BS to the $i$-th node for the $m$-th subchannel is described as
\begin{align}
    \mathbf{H}_{b,i,m}=U_{b,i} \textrm{PL}_{\textrm{LoS}}^{-\frac{1}{2}}\left(d_{b,i}\right)\mathbf{H}^{\textrm{LoS}}_{b,i,m}+\textrm{PL}_{\textrm{NLoS}}^{-\frac{1}{2}}\left(d_{b,i}\right)\mathbf{H}^{\textrm{NLoS}}_{b,i,m},
\end{align}
where $U_{b,i}$ is a Bernoulli random variable with LoS probability $P_{\textrm{LoS}}\left(d_{b,i}\right)$ and $d_{b,i}$ is the distance between the $b$-th BS and the $i$-th node.

\subsubsection{LoS probability model}
For each backhaul and access link, we consider the LoS probability model in \cite{channel_standard}.
In our IAB network scenario, the following three LoS probability cases are considered: 
\begin{align}\notag
    &\textrm{\textit{i)}~LoS~probability~for~backhaul~link~:}\\  &~~~P^\textrm{BH}_{\textrm{LoS}}(d)=\min\left(\frac{18}{d},1\right)\left(1-e^{-d/72}\right)+e^{-d/72},\\\notag
    &\textrm{\textit{ii)}~LoS~probability~for~access~link~from~MBS~to~UE~:}\\ 
    &~~~P^\textrm{MBS-UE}_{\textrm{LoS}}(d)=\min\left(\frac{18}{d},1\right)\left(1-e^{-d/63}\right)+e^{-d/63},\\\notag
    &\textrm{\textit{iii)}~LoS~probability~for~access~link~from~SBS~to~UE~:}\\ 
    &~~~P^\textrm{SBS-UE}_{\textrm{LoS}}(d)=0.5-\min(0.5, 5e^{-156/d})+\min(0.5, 5e^{-d/30}),
\end{align}
Based on the LoS probability model in (5)-(7), the existence of LoS path for each link is determined.
For the backhaul link, we adopt the cell site planning correction factor which can improve the LoS probability of backhaul by finding an optimal place among the candidate SBS sites \cite{channel_standard}.

\subsection{Signal Model}


IAB node is assumed to be operated in half-duplex mode, which is referred to as out-of-band full duplex (OBFD) \cite{OBFD}. 
In other words, the resources for parent backhaul link, child backhaul link, and access link are orthogonally allocated in frequency domain.
While the self-interference can be neglected due to the half-duplex mode \cite{Inter_type}, the cross-tier interference and the inter-cell interference should be considered in-between SBSs and the two types of interference considered in the IAB network scenario are illustrated in \cref{fig:system}.
Hence, the received signal of the $i$-th node from MBS, which is indexed by $0$, over the $m$-th subchannel is expressed as
\begin{align}\notag\label{1}
    r_{0,i,m}=&\sqrt{P_{0,m}}\mathbf{w}_{0,i,m}^H\mathbf{H}_{0,i,m}\mathbf{v}_{0,i,m}s_{0,i,m}\\\notag
    &+\!\!\underbrace{\sum_{b'\in \mathcal{B} \backslash \{i\}}\sum_{i'\in \mathcal{B}\backslash \{b',i\}}\sqrt{P_{b',m}}\mathbf{w}_{0,i,m}^H\mathbf{H}_{b',i,m}\mathbf{v}_{b',i',m}s_{b',i',m}}_\mathrm{Cross-tier~interference~by~backhaul~ link~between~SBSs}\\\notag
    &+\!\!\underbrace{\sum_{b'\in \mathcal{B} \backslash \{i\}}\sum_{i'\in \mathcal{K}\backslash\{i\}}\sqrt{P_{b',m}}\mathbf{w}_{0,i,m}^H\mathbf{H}_{b',i,m}\mathbf{v}_{b',i',m}s_{b',i',m}}_\mathrm{Cross-tier~interference~by~access~ link~in~small~cell}\\
    &+\mathbf{w}_{0,i,m}^H\mathbf{n}, ~\forall ~i\in\mathcal{I},~m\in\mathcal{M},
\end{align}
where $s_{b,i,m}$ is the transmitted signal from the $b$-th BS to the $i$-th node on the $m$-th subchannel with $E[|s_{b,i,m}|^2]=1$, $\mathbf{n}$ is the additive white gaussian noise (AWGN) with zero mean and covariance matrix, $\sigma_n^2\mathbf{I}$, and $P_{b,m}$ is the allocated power at the $b$-th BS over the $m$-th subchannel.
Note that $\mathbf{H}_{b,i,m}$ has the size of $N_s\times N_t$ for $i\in \mathcal{B}$ and ${N_u\times N_t}$ for $i \in \mathcal{K}$, in which $N_t=N_m$ for $b=0$ and $N_t=N_s$ for $b\in \mathcal{B}$.

Likewise, the received signal of the $i$-th node from the $b$-th BS on the $m$-th subchannel is expressed as
\begin{align}\notag\label{2}
    r_{b,i,m}\!\!\!&\\\notag
    =&\sqrt{P_{b,m}}\mathbf{w}_{b,i,m}^H\mathbf{H}_{b,i,m}\mathbf{v}_{b,i,m}s_{b,i,m}\\\notag
    &+\underbrace{\sum_{i'\in \mathcal{I}\backslash \{b\}}\sqrt{P_{0,m}}\mathbf{w}_{b,i,m}^H\mathbf{H}_{0,i,m}\mathbf{v}_{0,i',m}s_{0,i',m}}_\mathrm{Cross-tier~interference}\\\notag
    &+\!\!\underbrace{\sum_{b'\in \mathcal{B} \backslash \{b,i\}}\sum_{i'\in \mathcal{I} \backslash \{b,b',i\}}\sqrt{P_{b',m}}\mathbf{w}_{b,i,m}^H\mathbf{H}_{b',i,m}\mathbf{v}_{b',i',m}s_{b',i',m}}_\mathrm{Inter-cell~interference}\\
    &+\mathbf{w}_{b,i,m}^H\mathbf{n}, ~\forall~ b\in \mathcal{B},~ i\in\mathcal{I}\backslash\{b\}
\end{align}
We note that in \eqref{1} and \eqref{2}, the cross-tier interference and the inter-cell interference can be neglected if the subchannels are orthogonally allocated between all the wireless links. 

Let $x_{b,i,m}$ denotes the subchannel allocation variable for the $i$-th node served by the $b$-th BS over the $m$-th subchannel.
Then, the signal-to-interference-plus-noise ratio (SINR) is expressed by
\begin{align}\notag
    &\gamma_{b,i,m}(\mathbf{x},\mathbf{P})\\
    &~~~=\frac{\alpha_{b,i,m}P_{b,m}}{\sum_{b'\in \mathcal{B}_0 \backslash \{b,i\}}\sum_{i'\in \mathcal{I}\backslash \{b,b',i\}}x_{b',i',m}\alpha_{b,b',i,i',m}P_{b',m}+\sigma_n^2},
\end{align}
where $\alpha_{b,i,m}=|\mathbf{w}_{b,i,m}^H\mathbf{H}_{b,i,m}\mathbf{v}_{b,i,m}|^2$ and $\alpha_{b,b',i,i',m}=|\mathbf{w}_{b,i,m}^H\mathbf{H}_{b',i,m}\mathbf{v}_{b',i',m}|^2$.
Here, $\sigma_n^2$ is the noise power, and $\mathbf{x}$ and $\mathbf{P}$ are the vector whose elements are $x_{b,i,m}$ and $P_{b,m}$, respectively. 
Note that the interference occurs when other links use the same $m$-th subchannel, i.e., $x_{b',i',m}=1$, and interference does not exist, otherwise.
Thus, the achievable rate of the $i$-th node from the $b$-th BS on the $m$-th subchannel in bits per second (bps) is given by
\begin{align}
    R_{b,i,m}=W\log_2(1+\gamma_{b,i,m}),
\end{align}
where $W$ is the bandwidth per subchannel.

\section{Problem Formulation for Multi-Hop IAB Network}\label{problem}
In this section, we jointly formulate the problem of association optimization (AO), subchannel allocation (SA) and power allocation (PA) to maximize the sum rate of UEs for MH-IAB network.
Let $y_{b,i}$ denote the association variable where $y_{b,i}=1$ indicates that the $i$-th node is associated with the $b$-th BS and $y_{b,i}=0$, otherwise. 
For MH-IAB network scenario, the optimization problem can be formulated as
\begin{align}\notag\label{P1}
 (\textbf{P1})~\underset{\mathbf{y},\mathbf{x},\mathbf{P}}
{\max}&\sum_{b\in \mathcal{B}_0}\sum_{i\in \mathcal{K}}\sum_{m \in \mathcal{M}}y_{b,i}x_{b,i,m}R_{b,i,m}\\\notag
\mathrm{s.t}~&C_1:~\sum_{b\in \mathcal{B}_0}\sum_{m \in \mathcal{M}}y_{b,i}x_{b,i,m}R_{b,i,m}\geq R_i^\textrm{th}, ~\forall~ i\in \mathcal{K}\\\notag
~&C_2:~\sum_{b'\in \mathcal{B}_0}\sum_{m \in \mathcal{M}}y_{b',b}x_{b',b,m}R_{b',b,m}\\\notag  &~~~~~~~~~\geq \sum_{i\in \mathcal{I}}\sum_{m \in \mathcal{M}}y_{b,i}x_{b,i,m}R_{b,i,m},~\forall~b\in \mathcal{B}\\\notag
~&C_3:~\sum_{i\in \mathcal{I}}x_{b,i,m} \leq 1, ~\forall~ b\in \mathcal{B}_0, m\in \mathcal{M}\\\notag
~&C_4:~\sum_{i\in \mathcal{I}} x_{b,i,m}+\sum_{b'\in \mathcal{B}_0}x_{b',b,m} \leq 1,~ \forall ~b\in \mathcal{B},m\in \mathcal{M}\\\notag
~&C_5:~\sum_{b\in\mathcal{B}_0} y_{b,i}=1, ~\forall i\in \mathcal{I}\\\notag
~&C_6:~\sum_{b\in \mathcal{B}}y_{0,b}\geq 1\\\notag
~&C_7:~y_{b,b'}+y_{b',b}\leq 1,~\forall~b,b'\in \mathcal{B}_0\\\notag
~&C_8:~\sum_{m\in \mathcal{M} }P_{b,m} \leq P^\textrm{max}_b,~\forall~ b\in\mathcal{B}_0\\\notag
~&C_9:~P_{b,m} \geq 0,~\forall~b\in\mathcal{B}_0, m\in \mathcal{M} \\\notag
~&C_{10}:~y_{b,i}  \in \{0,1\}, ~\forall~b\in\mathcal{B}_0, i\in \mathcal{I}\\
~&C_{11}:~x_{b,i,m} \in \{0,1\}, ~\forall~b\in\mathcal{B}_0, i\in \mathcal{I}, m\in \mathcal{M}
\end{align}
where 
$\mathbf{y}$ is the vector whose elements are $y_{b,i}$, $R_i^\mathrm{th}$ is the minimum data rate requirement of the $i$-th UE, and $P^\mathrm{max}_b$ is the maximum transmit power at the $b$-th BS.
The constraint $C_1$ in \eqref{P1} is the QoS requirement for each UE. In $C_2$, the data rate of parent backhaul link must be no less than the sum rate of all the access and child backhaul links served by the $b$-th BS.
For the constraints related to subchannel, $C_3$ indicates that each subchannel is allocated to at most one child link (i.e., access or child backhaul link) served by the $b$-th BS, while $C_4$ represents the half-duplex mode of IAB node. $C_5$ ensures that each node is associated with only one BS. Since MBS is only connected to the core networks, at least one IAB node should be connected to MBS as described in $C_6$. Also, $C_7$ ensures that all the backhaul links are directional.
$C_8$ and $C_9$ denote the transmit power constraint for each BS and the non-negative transmit power, respectively. Finally, $C_{10}$ and $C_{11}$ indicate that both association and SA variables are binary.
Note that without loss of generality, $y_{b,b}=0$ and $x_{b,b,m}=0$ are assumed for $b\in \mathcal{B}_0$ since the same transceiver cannot be connected.

Due to binary and continuous variables, \eqref{P1} is regarded as a mixed integer non-linear programming (MINLP) which is a non-convex and NP-hard problem in general.
To solve it, an exhaustive search is required but it is infeasible in practice due to the prohibitive complexity.
Several algorithms such as Branch-and-bound can be used to obtain the optimal solution of MINLP problem. However, its complexity for the worst case is as high as an exhaustive search \cite{MIP}. 
 Therefore, we propose the suboptimal algorithm with low complexity by decomposing the original problem into three sub-problems.


\section{Joint Association Optimization and Resource Allocation for Multi-Hop IAB Network}\label{solution}
To address the non-convex problem in \eqref{P1}, we propose an alternating optimization algorithm.
By decoupling the original problem into subproblems and iteratively updating solutions from each subproblem, the original problem can be efficiently solved.
In the following subsections, we present the three subproblems for each AO, SA, and PA, and then introduce an alternating algorithm to solve these three subproblems. 



\subsection{Association Optimization for Bakchaul and Access Links}
For given SA variable $\mathbf{x}$ and PA variable $\mathbf{P}$, \eqref{P1} can be transformed into the association problem given as
\begin{align}\label{P2}
    (\textbf{P2})~\underset{\mathbf{y}}{\max}& \sum_{b\in \mathcal{B}_0}\sum_{i\in \mathcal{K}}y_{b,i}\hat{R}_{b,i},~\mathrm{s.t}~C_1, C_2, C_5, C_6, C_7, C_{10},
\end{align}
where $\hat{R}_{b,i}=\sum_{m \in \mathcal{M}}x_{b,i,m}R_{b,i,m}$.
However, \eqref{P2} is still  non-convex problem due to the binary constraint $C_{10}$.
To make the problem more tractable, we relax $y_{b,i}$ as a continuous real variable with the range of [0, 1], where the relaxation provides a zero-duality gap \cite{time_sharing}.
In that case, $y_{b,i}$ can be regarded as a timing sharing factor which is the fraction of time when the $i$-th node is associated with the $b$-th BS.
By relaxing the binary variable into the continuous variable, \eqref{P2} is a linear programming which can be solved using dual problem due to a zero duality gap.
The Lagrangian function of \eqref{P2} can be defined as 
\begin{align}\notag\label{lag}
    L(\mathbf{y},\boldsymbol{\lambda},\boldsymbol{\mu},\nu)= &\sum_{b\in \mathcal{B}_0}\sum_{i\in \mathcal{K}}y_{b,i}\hat{R}_{b,i}\\\notag
    &+\sum_{i\in \mathcal{K}}\lambda_i\left(\sum_{b\in \mathcal{B}_0}y_{b,i}\hat{R}_{b,i}- R_i^\mathrm{th}\right)\\\notag
    &+\sum_{b\in \mathcal{B}}\mu_b\left(\sum_{b'\in \mathcal{B}_0} y_{b',b}\hat{R}_{b',b}-\sum_{i\in \mathcal{I}}y_{b,i}\hat{R}_{b,i}\right)\\
    &+\nu\left(\sum_{b\in \mathcal{B}}y_{0,b}-1\right),
\end{align}
where $\boldsymbol{\lambda}=[\lambda_{B+1},\dots,\lambda_{B+K}]^T$, $\boldsymbol{\mu}=[\mu_{1},\dots,\mu_{B}]^T$, and $\nu$ are the dual variables for the constraints $C_1$, $C_2$, and $C_6$, respectively.
Then, the Lagrangian dual problem of \eqref{P2} can be formulated as
\begin{align}\label{dual_problem}
  \underset{\boldsymbol{\lambda},\boldsymbol{\mu},\nu}{\min}~ g(\boldsymbol{\lambda},\boldsymbol{\mu},\nu),~
  \mathrm{s.t}~\boldsymbol{\lambda},\boldsymbol{\mu},\nu \geq 0
\end{align}
In \eqref{dual_problem}, the dual function $g(\lambda,\mu,\nu)$ can be obtained as
\begin{align}\notag\label{dual_function}
  g(\boldsymbol{\lambda},\boldsymbol{\mu},\nu)=\underset{\mathbf{y}}{\max}&~ L(\mathbf{y},\boldsymbol{\lambda},\boldsymbol{\mu},\nu)\\
  \mathrm{s.t}~ &C_5, C_7,~0\leq y_{b,i} \leq 1,~\forall~b\in\mathcal{B}_0, i\in \mathcal{I}
\end{align}
The Lagrangian dual problem can be solved by decomposing it into inner and outer problems.
We first solve the inner problem in \eqref{dual_function} to obtain the optimal $y_{b,i}$ for given dual variables and the dual variables are updated at the outer problem in \eqref{dual_problem}.
As a result, the inner and outer problems can be iteratively solved until convergence.

For the inner problem, the Lagrangian function can be rewritten as
\begin{align}\label{lag_re}
    L&(\mathbf{y},\boldsymbol{\lambda},\boldsymbol{\mu},\nu) = \sum_{b\in \mathcal{B}_0}\sum_{i\in \mathcal{I}}X_{b,i}y_{b,i}
\end{align}
where
\begin{align}
    X_{b,i}=\left\{\begin{matrix*}[l]
    (1+\lambda_i)\hat{R}_{0,i}, & \mathrm{for}~ b=0, i\in \mathcal{K} \\
    (1+\lambda_i-\mu_b)\hat{R}_{b,i}, & \mathrm{for}~ b\in \mathcal{B}, i\in \mathcal{K}\\
    \mu_i\hat{R}_{0,i}+\nu, & \mathrm{for}~ b=0, i\in \mathcal{B}\\
    (\mu_i-\mu_b)\hat{R}_{b,i}, & \mathrm{for}~ b,i\in \mathcal{B}
    \end{matrix*}\right.
\end{align}
Since the $i$-th node needs to select only one BS as in the constraint $C_5$, the optimal solution of \eqref{dual_function} can be obtained as
\begin{align}\label{UA_sol}
    y_{b,i}=\left\{\begin{matrix*}[l]
    1, & b^*=\underset{b}{\arg}\max X_{b,i} ~\forall ~i\in \mathcal{I}\\
    0, & \mathrm{otherwise}
    \end{matrix*}\right.
\end{align}
Since $X_{b',b}=(\mu_{b'}-\mu_b)\hat{R}_{b',b}$ and $X_{b,b'}=(\mu_b-\mu_{b'})\hat{R}_{b,b'}$ have a different sign, both $y_{b,b'}$ and $y_{b',b}$ cannot be 1 at the same time which makes the solution in \eqref{UA_sol} satisfy the constraint $C_7$.

To update the dual variables at the outer problem, we adopt the subgradient method given as \cite{subgradient}
\begin{align}\label{dual_var1}
    &\lambda_i^{(t+1)}=\left[\lambda_i^{(t)}-\delta_1^{(t)}\left(\sum_{b\in \mathcal{B}_0}y_{b,i}\hat{R}_{b,i}- R_i^{th}\right)\right]^+,\\\label{dual_var2}
    &\mu_b^{(t+1)}=\left[\mu_b^{(t)}-\delta_2^{(t)}\left(\sum_{b'\in \mathcal{B}_0} y_{b',b}\hat{R}_{b',b}-\sum_{i\in \mathcal{I}}y_{b,i}\hat{R}_{b,i}\right)\right]^+,\\\label{dual_var3}
    &\nu^{(t+1)}=\left[\nu^{(t)}-\delta_3^{(t)}\left(\sum_{b\in \mathcal{B}}y_{0,b}-1\right)\right]^+,
\end{align}
where $[x]^+=\max(x,0)$ and $\delta_1^{(t)}$, $\delta_2^{(t)}$, and $\delta_3^{(t)}$ are the positive step sizes at the $t$-th iteration\footnote{The step size is typically chosen as the constant or square summable but not summable that satisfies $\sum_{k=1}^{\infty}\delta^{(t)^2}<\infty$ and $\sum_{k=1}^{\infty}\delta^{(t)}=\infty$. One example is $\delta^{(t)}=\frac{a}{b+t}$ where $a>0$ and $b\geq 0$ \cite{subgradient}.}.
When using the appropriate step size, the subgradient method guarantees to converge to the optimal solution since the dual problem is always convex problem.

Based on the solutions of \eqref{UA_sol}-\eqref{dual_var3}, we propose an iterative algorithm to obtain the association of backhaul and access links. Setting up the maximum iteration number $T_\mathrm{max}$, the proposed algorithm is iteratively continued  until convergence or $t=T_{max}$.
In each iteration, the optimal association result is obtained for given dual variables, and dual variables are updated using the subgradient method.
Therefore, the algorithm of link association can be summarized as in Algorithm 1.

\begin{algorithm}
\caption{Lagrangian-based AO Algorithm}\label{alg1}

\begin{algorithmic}[1]
\STATE Initialize $T_\mathrm{max}$, dual variables $\boldsymbol{\lambda},\boldsymbol{\mu},\nu$ and feasible $x_{b,i,m}$ and $P_{b,m}$
\STATE Set $t=0$
\REPEAT 
\STATE Obtain the association policy $y_{b,i}^*$ as in \eqref{UA_sol}
  \STATE Update $\lambda^{(t+1)}$, $\mu^{(t+1)}$, and $\nu^{(t+1)}$ as in \eqref{dual_var1}-\eqref{dual_var3}
  \STATE Set $t=t+1$
\UNTIL{convergence or $t=T_\mathrm{max}$}
\end{algorithmic}
\end{algorithm}




\subsection{Subchannel Allocation}
In this subsection, we optimize SA given AO variable $\mathbf{y}$ and PA variable $\mathbf{P}$.
The subproblem for SA can be written as
\begin{align}\notag\label{P3}
        (\textbf{P3})~\underset{\mathbf{x}}{\max}& \sum_{b\in \mathcal{B}_0}\sum_{i\in \mathcal{K}}\sum_{m \in \mathcal{M}}y_{b,i}x_{b,i,m}R_{b,i,m}\\
\mathrm{s.t}~&C_1, C_2, C_3, C_4, C_{11}
\end{align}
Note that \eqref{P3} is a non-convex problem due to the binary constraint $C_{11}$ and the non-convex function of $x_{b,i,m}R_{b,i,m}$, which is generally difficult to solve.
To handle the non-convexity, we adopt a low complexity iterative algorithm based on successive convex approximation (SCA) that approximates the non-convex problem to a sequence of convex problem. It is known that SCA method guarantees the convergence to a local optimum \cite{SCA}.

First, we need to approximate the non-convex function $x_{b,i,m}R_{b,i,m}$ with the following inequality that is based on first order Taylor series \cite{inequality}.
\begin{align}\label{inequal}
    v\ln(1+\frac{1}{z})\geq (2-\frac{\Bar{v}}{v})\Bar{v}\ln(1+\frac{1}{\Bar{z}})+\frac{\Bar{v}}{\Bar{z}+1}(1-\frac{z}{\Bar{z}}).
\end{align}
By substituting $v=x_{b,i,m}$, $\Bar{v}=x_{b,i,m}^{(t)}$, $z=1/\gamma_{b,i,m}(\mathbf{x},\mathbf{P})$, and $\Bar{z}=1/\gamma_{b,i,m}(\mathbf{x}^{(t)},\mathbf{P})$  into \eqref{inequal} for given the obtained SA variable at the $t$-th iteration, denoted by $\mathbf{x}^{(t)}$, we approximate the function $x_{b,i,m}R_{b,i,m}$, and then obtain its lower bound as \eqref{rate_approx}.
\begin{figure*}[t]
\begin{align}\notag\label{rate_approx}
    x_{b,i,m}R_{b,i,m}&\geq \bar{R}_{b,i,m}(\mathbf{x}^{(t)})\\\notag
    &=\frac{x_{b,i,m}^{(t)}}{\ln2}\left(2-\frac{x_{b,i,m}^{(t)}}{x_{b,i,m}}\right)\log_2\left(1+\frac{\alpha_{b,i,m}P_{b,m}}{\sum_{b'\in \mathcal{B}_0 \backslash \{b\}}\sum_{i'\in \mathcal{K}\backslash \{b\}}x_{b',i',m}P_{b',m}\alpha_{b,b',i,i',m}+\sigma^2}\right)\\\notag
    &~~+\frac{\alpha_{b,i,m}P_{b,m}x_{b,i,m}^{(t)}/\ln2}{\alpha_{b,i,m}P_{b,m}+\sum\limits_{b'\in \mathcal{B}_0 \backslash \{b\}}\sum\limits_{i'\in \mathcal{K}\backslash \{b\}}x_{b',i',m}^{(t)}P_{b',m}\alpha_{b,b',i,i',m}+\sigma^2}\left(1-\frac{\sum\limits_{b'\in \mathcal{B}_0 \backslash \{b\}}\sum\limits_{i'\in \mathcal{K}\backslash \{b\}}x_{b',i',m}P_{b',m}\alpha_{b,b',i,i',m}+\sigma^2}{\sum\limits_{b'\in \mathcal{B}_0 \backslash \{b\}}\sum\limits_{i'\in \mathcal{K}\backslash \{b\}}x_{b',i',m}^{(t)}P_{b',m}\alpha_{b,b',i,i',m}+\sigma^2}\right).\\
\end{align}
\hrule
\end{figure*}
It is obvious that $\bar{R}_{b,i,m}(\mathbf{x}^{(t)})$ in \eqref{rate_approx} is a concave function with respect to $x_{b,i,m}$.

\emph{Proposition 1: The approximation of \eqref{rate_approx} provides a tight lower bound.}

\emph{Proof:} Due to the inequality in \eqref{inequal}, $x_{b,i,m}R_{b,i,m}\geq \bar{R}_{b,i,m}(\mathbf{x}^{(t)})$ holds as in \eqref{rate_approx}. Note that its equality holds for $\mathbf{x}=\mathbf{x}^{(t)}$ which shows the tightness of the lower bound.

Using the approximated concave function $\bar{R}_{b,i,m}(\mathbf{x}^{(t)})$, \eqref{P3} can be reformulated as
\begin{align}\notag\label{P3.1}
 (\textbf{P3-1})~\underset{\mathbf{x}}{\max}& \sum_{b\in \mathcal{B}_0}\sum_{i\in \mathcal{K}}\sum_{m \in \mathcal{M}}y_{b,i}\bar{R}_{b,i,m}(\mathbf{x}^{(t)})\\\notag
\mathrm{s.t}~&\bar{C}_1:~\sum_{b\in \mathcal{B}_0}\sum_{m \in \mathcal{M}}y_{b,i}\bar{R}_{b,i,m}(\mathbf{x}^{(t)})\geq R_i^\mathrm{th}\\\notag
~&\bar{C}_2:~\sum_{m \in \mathcal{M}}\sum_{b'\in \mathcal{B}_0}y_{b',b}\bar{R}_{b',b,m}(\mathbf{x}^{(t)})\\\notag
&~~~~~~~-\sum_{m \in \mathcal{M}}\sum_{i\in \mathcal{I}}y_{b,i}\bar{R}_{b,i,m}(\mathbf{x}^{(t)})\geq 0\\
&C_3, C_4, C_{11}
\end{align}
Note that \eqref{P3.1} provides a tight lower bound as \emph{Proposition 1}. Furthermore, it converges to KKT point of the original problem using an iterative algorithm since it provides a sequence of non-decreasing objective function value \cite{KKT1,KKT2}.
Therefore, we focus on the lower bound of original problem.


However, \eqref{P3.1} is still non-convex problem since it is the integer programming and $\bar{C}_{2}$ is non-convex due to the difference of two concave functions.
To tackle the non-convexity of binary constraint first, we rewrite the constraint $C_{11}$ into its equivalent form given as
\begin{align}\notag\label{int_relax}
    &C_{11a}:~0\leq x_{b,i,m}\leq 1,~\forall~b\in\mathcal{B}_0,~ i\in \mathcal{I},~m\in\mathcal{M}\\
    &C_{11b}:~\sum_{b\in \mathcal{B}_0}\sum_{i\in \mathcal{I}}\sum_{m \in \mathcal{M}}(x_{b,i,m}-x_{b,i,m}^2)\leq0
\end{align}
It should be noted that $C_{11b}$ is non-convex constraint.
Hence, we adopt the Lagrangian approach by adding a penalty term to the objective function \cite{penalty1,penalty2}.
Accordingly, \eqref{P3.1} can be rewritten as
\begin{align}\notag\label{P3.2}
 \underset{\mathbf{x}}{\max}& \sum_{b\in \mathcal{B}_0}\sum_{m \in \mathcal{M}}\left(\sum_{i\in \mathcal{K}}y_{b,i}\bar{R}_{b,i,m}(\mathbf{x}^{(t)})-\mu \sum_{i\in \mathcal{I}}(x_{b,i,m}-x_{b,i,m}^2)\right)\\
\mathrm{s.t}~&\bar{C}_1, \bar{C}_2, C_3, C_4, C_{11a}
\end{align}
where $\mu$ is the penalty factor which penalizes the objective function for non-integer value of $x_{b,i,m}$.

\emph{Proposition 2: For sufficiently large value of $\mu$, the optimization problem \eqref{P3.2} is equivalent to  \eqref{P3.1}.}

\emph{Proof: See Appendix A}

Since the objective function and $\bar{C}_2$ are still non-convex functions, we can rewrite \eqref{P3.2} as
\begin{align}\label{P3.3}\notag
    (\textbf{P3-2})~\underset{\mathbf{x}}{\max}&~F(\mathbf{x})-G(\mathbf{x})\\\notag
\mathrm{s.t}~&f_b(\mathbf{x})-g_b(\mathbf{x})\geq 0,~ \forall~ b\in\mathcal{B}\\
&\bar{C_1}, C_3, C_4, C_{11a}
\end{align}
where $F(\mathbf{y})$, $G(\mathbf{x})$, $f_b(\mathbf{x})$, and $g_b(\mathbf{x})$ are defined as
\begin{align}\notag
&F(\mathbf{x})=\sum\limits_{b\in \mathcal{B}_0}\sum\limits_{m \in \mathcal{M}}\left(\sum\limits_{i\in \mathcal{K}}y_{b,i}\bar{R}_{b,i,m}(\mathbf{x}^{(t)})-\mu \sum\limits_{i\in \mathcal{I}}x_{b,i,m}\right),\\\notag
&G(\mathbf{x})=-\mu \sum\limits_{b\in \mathcal{B}_0}\sum\limits_{i\in \mathcal{I}}\sum\limits_{m \in \mathcal{M}}x_{b,i,m}^2,\\ \notag
&f_b(\mathbf{x})=\sum\limits_{b'\in \mathcal{B}_0}\sum\limits_{m \in \mathcal{M}}y_{b',b}\bar{R}_{b',b,m}(\mathbf{x}^{(t)}),\\
&g_b(\mathbf{x})=\sum\limits_{i\in \mathcal{I}}\sum\limits_{m \in \mathcal{M}}y_{b,i}\bar{R}_{b,i,m}(\mathbf{x}^{(t)}).
\end{align}
It can be observed that \eqref{P3.3} is a form of the difference of two concave functions (DC) \cite{DCA}.
Since $F(\mathbf{x})$, $G(\mathbf{x})$, $f_b(\mathbf{x})$, and $g_b(\mathbf{x})$ are concave functions, \eqref{P3.3} belongs to the class of DC programming (DCP) which can be solved using DC algorithm (DCA) \cite{DCA}.
Thus, we can construct a surrogate function for $G(\mathbf{x})$ and $g_b(\mathbf{x})$ by using the first order Taylor approximation given as
\begin{align}\notag\label{approx}
    &G(\mathbf{x})\leq \bar{G}(\mathbf{x},\mathbf{x}^{(t)})= G(\mathbf{x}^{(t)})+\nabla_\mathbf{x} G^T(\mathbf{x}^{(t)})(\mathbf{x}-\mathbf{x}^{(t)}).\\
   &g_b(\mathbf{x})\leq \bar{g}_b(\mathbf{x},\mathbf{x}^{(t)})= g_b(\mathbf{x}^{(t)})+\nabla_\mathbf{x} g_b^T(\mathbf{x}^{(t)})(\mathbf{x}-\mathbf{x}^{(t)}).
\end{align}
As a result, the optimal SA can be obtained by solving the following convex problem.
\begin{align}\label{P3.4}\notag
    (\textbf{P3-3})~\underset{\mathbf{x}}{\max}&~F(\mathbf{x})-\bar{G}(\mathbf{x},\mathbf{x}^{(t)})\\
\mathrm{s.t}~&f_b(\mathbf{x})-\bar{g}_b(\mathbf{x},\mathbf{x}^{(t)})\geq 0,~\bar{C_1}, C_3, C_4, C_{11a}
\end{align}
This problem can be solved using the conventional optimization method such as interior point method \cite{cvx_program,CVX}.


\emph{Proposition 3: The optimal solution of \eqref{P3.4} is the subset of the feasible set of \eqref{P3.3}}

\emph{Proof:} In the $(t+1)$-th iteration, the solution obtained from \eqref{P3.4} is $\mathbf{x}^{(t+1)}$. It is noteworthy that $f_b(\mathbf{x})-g_b(\mathbf{x})\geq f_b(\mathbf{x})-\bar{g}_b(\mathbf{x},\mathbf{x}^{(t)})$ as in \eqref{approx}.
Therefore, the constraint $f_b(\mathbf{x}^{(t+1)})-\bar{g}_b(\mathbf{x}^{(t+1)},\mathbf{x}^{(t)})\geq 0$ in \eqref{P3.4} guarantees $f_b(\mathbf{x}^{(t+1)})-g_b(\mathbf{x}^{(t+1)})\geq 0$ of \eqref{P3.3}.

\emph{Proposition 4: The approximation of \eqref{approx} gives a tight lower bound  which provides the improved solution after each iteration and it converges to a local optimal point.}

\emph{Proof:} Recall that the objective function of \eqref{P3.3} can be approximated as $F(\mathbf{x})-G(\mathbf{x})\geq F(\mathbf{x})-\bar{G}(\mathbf{x},\mathbf{x}^{(t)})$. For $\mathbf{x}=\mathbf{x}^{(t)}$, the equality holds, which means the tightness of the lower bound. Moreover, the solution of \eqref{P3.4} gives the improved objective value at each iteration. For given $\mathbf{x}^{(t)}$, the optimal solution of \eqref{P3.3} at the $(t+1)$-th iteration is $\mathbf{x}^{(t+1)}$. Then, we have
\begin{align}\notag
    F&(\mathbf{x}^{(t+1)})-G(\mathbf{{x}}^{(t+1)})\\\notag
    &\geq F(\mathbf{x}^{(t+1)})-G(\mathbf{x}^{(t)})-\nabla_\mathbf{x} G^T(\mathbf{x}^{(t+1)})(\mathbf{x}^{(t+1)}-\mathbf{x}^{(t)})\\\notag
    &=\underset{\mathbf{x}}{\max}~ F(\mathbf{x})-G(\mathbf{x}^{(t)})-\nabla_\mathbf{x} G^T(\mathbf{x})(\mathbf{x}-\mathbf{x}^{(t)})\\\notag
    &\geq F(\mathbf{x}^{(t)})-G(\mathbf{x}^{(t)})-\nabla_\mathbf{x} G^T(\mathbf{x}^{(t)})(\mathbf{x}^{(t)}-\mathbf{x}^{(t)})\\
    &\geq F(\mathbf{x}^{(t)})-G(\mathbf{x}^{(t)}).
\end{align}
Thus, the solution of \eqref{P3.4} provides the non-decreasing objective value as the iteration continues and it converges to a local optimal solution.

It is worth noting that the lower bound will become tighter as the iteration continues.
To tighten the obtained lower bound, we adopt the iterative algorithm which continues until convergence to a local optimum point.
The detailed procedure is summarized in Algorithm 2.

\begin{algorithm}
\caption{DCA-based iterative algorithm for SA}\label{alg2}

\begin{algorithmic}[1]
\STATE  Initialize $T_\mathrm{max}$, penalty factor $\mu$, and feasible point $\mathbf{x}^{(0)}$, $\mathbf{y}$ and $\mathbf{P}$
\STATE Set $t=0$
\REPEAT 
\STATE Calcluate $\bar{R}_{b,i,m}(\mathbf{x}^{(t)})$, $\bar{G}(\mathbf{x},\mathbf{x}^{(t)})$, and $\bar{g}_b(\mathbf{x},\mathbf{x}^{(t)})$
\STATE Solve the problem \eqref{P3.4} and obtain the subchannel allcoation policy $\{\mathbf{x}^{(t)^*}\}$
\STATE Set $t=t+1$
\UNTIL{convergence or $t=T_\mathrm{max}$}
\end{algorithmic}
\end{algorithm}




\subsection{Power Allocation}
For given AO variable $\mathbf{y}$ and SA variable $\mathbf{x}$, the PA problem can be reformulated as 
\begin{align}\label{P4}\notag
    (\textbf{P4})~\underset{\mathbf{P}}{\max}&~ \sum_{b\in \mathcal{B}_0}\sum_{i\in \mathcal{K}}\sum_{m \in \mathcal{M}}y_{b,i}x_{b,i,m}R_{b,i,m}\\\mathrm{s.t}~&C_1, C_2, C_8, C_9
\end{align}
Our aim for this subproblem is to obtain the optimal PA that maximizes the sum rate under the constraints in \eqref{P4}.
It is well-known that the water-filling algorithm is optimal PA method. However, we are not able to apply the conventional water-filling algorithm to our system model since there exist two kinds of interference (e.g., the cross-tier interference and the inter-cell interference) in IAB network.
Therefore, we propose the iterative algorithm in consideration of cross-tier interference and inter-cell interference.

Note that \eqref{P4} is a non-convex problem due to the non-convexity of the objective function and feasible set.
To make the problem tractable, we reformulate the problem \eqref{P4} into the form of DCP which can be solved by DCA.
We first rewrite the objective function as
\begin{align}\label{dc_rate}
   y_{b,i}x_{b,i,m}R_{b,i,m}=e_{b,i,m}(\mathbf{P})-q_{b,i,m}(\mathbf{P}),
\end{align}
where $e_{b,i,m}(\mathbf{P})$ and $q_{b,i,m}(\mathbf{P})$ is defined as \eqref{dc_notation} at the top of the current page.
\begin{figure*}[t]
\begin{align}\notag\label{dc_notation}
    &e_{b,i,m}(\mathbf{P})=y_{b,i}x_{b,i,m}\log_2\left(\alpha_{b,i,m}P_{b,m}+\sum\limits_{b'\in \mathcal{B}_0 \backslash \{b,i\}}\sum\limits_{i'\in \mathcal{I}\backslash \{b,b',i\}}x_{b',i',m}\alpha_{b,b',i,i',m}P_{b',m}+\sigma^2\right).\\
        & q_{b,i,m}(\mathbf{P})=y_{b,i}x_{b,i,m}\log_2\left(\sum\limits_{b'\in \mathcal{B}_0 \backslash \{b,i\}}\sum\limits_{i'\in \mathcal{K}\backslash \{b,b',i\}}x_{b',i',m}\alpha_{b,b',i,i',m}P_{b',m}+\sigma^2\right).
\end{align}
\hrule
\end{figure*}
Using \eqref{dc_rate}, \eqref{P4} can be written in the form of DC as follows.
\begin{align}\notag\label{P4.1}
    (\textbf{P4-1})~\underset{\mathbf{P}}{\max}&~ E(\mathbf{P})-Q(\mathbf{P})\\\notag
~\mathrm{s.t}~&E_{i}(\mathbf{P})-Q_{i}(\mathbf{P})\geq R_i^\mathrm{th},~\forall~i\in\mathcal{K}\\\notag
&h_{1,b}(\mathbf{P})-h_{2,b}(\mathbf{P})\geq 0, ~\forall~b\in\mathcal{B}\\
&C_8, C_9
\end{align}
where
\begin{align}\notag\label{dc_abbrev}
    &E(\mathbf{P})=\sum_{b\in \mathcal{B}_0}\sum_{i\in \mathcal{K}}\sum_{m \in \mathcal{M}}e_{b,i,m}(\mathbf{P}),\\\notag
    &Q(\mathbf{P})=\sum_{b\in \mathcal{B}_0}\sum_{i\in \mathcal{K}}\sum_{m \in \mathcal{M}}q_{b,i,m}(\mathbf{P}),\\\notag
    &E_{i}(\mathbf{P})=\sum\limits_{b\in \mathcal{B}_0}\sum\limits_{m \in \mathcal{M}}e_{b,i,m}(\mathbf{P}),\\\notag
    &Q_{i}(\mathbf{P})=\sum\limits_{b\in \mathcal{B}_0}\sum\limits_{m \in \mathcal{M}}q_{b,i,m}(\mathbf{P}),\\\notag
    &h_{1,b}(\mathbf{P})=\sum_{b'\in \mathcal{B}_0}\sum_{m \in \mathcal{M}}e_{b',b,m}(\mathbf{P})+\sum_{i\in \mathcal{I}}\sum_{m \in \mathcal{M}}q_{b,i,m}(\mathbf{P}),\\
    &h_{2,b}(\mathbf{P})=\sum_{b'\in \mathcal{B}_0}\sum_{m \in \mathcal{M}}q_{b',b,m}(\mathbf{P})+\sum_{i\in \mathcal{I}}\sum_{m \in \mathcal{M}}e_{b,i,m}(\mathbf{P}).
\end{align}
Since $e_{b,i,m}(\mathbf{P})$ and $q_{b,i,m}(\mathbf{P})$ are concave functions, all the functions in \eqref{dc_abbrev} are concave and \eqref{P4.1} can be considered as DCP.
We approximate $Q(\mathbf{P})$, $Q_{i}(\mathbf{P})$, and $h_{2,b}(\mathbf{P})$ using the first order Taylor series around the point $\mathbf{P}^{(t)}$ given as
\begin{align}\label{p_taylor}\notag
    &Q(\mathbf{P})\leq\bar{Q}(\mathbf{P},\mathbf{P}^{(t)})= Q(\mathbf{P}^{(t)})+\nabla_\mathbf{P} Q^T(\mathbf{P}^{(t)})(\mathbf{P}-\mathbf{P}^{(t)}),\\\notag
    &Q_{i}(\mathbf{P}) \leq \bar{Q}_i(\mathbf{P},\mathbf{P}^{(t)})=Q_{i}(\mathbf{P}^{(t)})+\nabla_\mathbf{P} Q_{i}^T(\mathbf{P}^{(t)})(\mathbf{P}-\mathbf{P}^{(t)}),\\
    &h_{2,b}(\mathbf{P}) \leq\bar{h}_{2,b}(\mathbf{P},\mathbf{P}^{(t)})= h_{2,b}(\mathbf{P}^{(t)})+\nabla_\mathbf{P} h_{2,b}^T(\mathbf{P}^{(t)})(\mathbf{P}-\mathbf{P}^{(t)}).
\end{align}
Then, the approximated problem at the $t$-th iteration can be written as
\begin{align}\notag\label{P4.2}
(\textbf{P4-2})~\underset{\mathbf{P}}{\max}&~E(\mathbf{P})-\bar{Q}(\mathbf{P},\mathbf{P}^{(t)})\\\notag
\mathrm{s.t}~&E_{i}(\mathbf{P})-\bar{Q}_{i}(\mathbf{P},\mathbf{P}^{(t)})\geq R_i^\mathrm{th}\\\notag
&h_{1,b}(\mathbf{P})-\bar{h}_{2,b}(\mathbf{P},\mathbf{P}^{(t)})\geq 0\\
&C_8, C_9
\end{align}
Note that \eqref{P4.2} is a convex problem which can be solved by conventional convex approach such as interior-point method. According to \emph{Proposition 3}, the feasible points of \eqref{P4.2} are the subset of the feasible points of \eqref{P4}.
It should be also noted that the solution of \eqref{P4.2}, which is obtained by the iterative algorithm similar to Algorithm 2, converges to a KKT point of the original problem given in \eqref{P4} \cite{multi_assoc}.



\subsection{Overall Algorithm}
In the previous subsections, we proposed the low complexity algorithms for AO, SA, and PA. The original problem in \eqref{P1} is decomposed into the three subproblems of \eqref{P2}, \eqref{P3}, and \eqref{P4} where the Lagrangian approach based algorithm is proposed for AO subproblem, and DCA based iteration algorithms are proposed for SA and PA subproblems.
We also showed that the optimal solution for AO can be obtained and the local optimal solution of the SA and PA can be achieved.
Now, we propose an alternating algorithm which jointly optimizes AO, SA, and PA by solving each subproblem in an iterative manner. 
The detail of the overall algorithm is summarized in Algorithm 3 and the convergence behavior of proposed algorithm is shown in Section \ref{simul}.


\begin{algorithm}
\caption{Joint AO, SA, and PA algorithm for MH-IAB network }\label{alg3}
\begin{algorithmic}[1]
\STATE Initialize $t=0$, $T_\mathrm{max}$ and feasible  $\mathbf{y}^{(0)}$, $\mathbf{x}^{(0)}$, and $\mathbf{P}^{(0)}$.
\REPEAT
\STATE Set $t=t+1$
\STATE Obtain the optimal $\mathbf{y}^{(t+1)}$ using \eqref{UA_sol} with given $\mathbf{P}^{(t)}$ and $\mathbf{x}^{(t)}$.
\STATE Solve the problem \eqref{P3.4} for given $\mathbf{y}^{(t+1)}$ and $\mathbf{P}^{(t)}$ to obtain $\mathbf{x}^{(t+1)}$
\STATE Solve the problem \eqref{P4.2} for given $\mathbf{y}^{(t+1)}$ and $\mathbf{x}^{(t+1)}$ and obtain the optimal  $\mathbf{P}^{(t+1)}$
\UNTIL{convergence or $t=T_\mathrm{max}$}
\end{algorithmic}
\label{Algorithm_Main}
\end{algorithm}

\subsection{Complexity Analysis}\label{sec:complexity}
The computational complexity of the block coordinate descent-based approach in Algorithm \ref{Algorithm_Main} lies mainly in Algorithm \ref{alg1} and Algorithm \ref{alg2}.
For Algorithm \ref{alg1}, considering that the total iterative number in Algorithm \ref{alg1} is $I_{1}$, the computational complexity of AO is given by $O( I_{1} (BK+K) )$.
For Algorithm \ref{alg2}, it involves solving the convex problem given in \eqref{P3.4} where the dimension of optimization variable is $B^{2}M + (K+1)BM +KM$.
By using a primal-dual interior point method to address the convex problem, the computational complexity of SA  is $O \Big( I_{2} (B^{2}M + (K+1)BM +KM) \log{\epsilon^{-1}} \Big)$, in which $I_{2}$ and $\epsilon$ represent the number of iteration in Algorithm \ref{alg2} and the determined accepted duality gap, respectively \cite{CVX}.
Likewise, the computational complexity of PA is given by $O \Big(I_{3} (BM + M) \log{\epsilon^{-1}} \Big)$, where $I_{3}$ is the number of iteration for \eqref{P4.2}.
Thus, given the total iteration number is $I$, the computational complexity of Algorithm \ref{Algorithm_Main} is derived as \eqref{complexity}.
\begin{figure*}[t]
\begin{align}
    O \Bigg( I I_{1} \big( BK+K \big) + I \Big( (I_{2} \big(B^{2}M + (K+1)BM +KM \big) + I_{3} (BM + M) \Big) \log{\epsilon^{-1}}  \Bigg). \label{complexity}
\end{align}
\hrule
\end{figure*}

\section{Simulation Results}\label{simul}

\begin{table}
\begin{center}
\caption{System Parameters}\label{Simul_table}
\label{table_notation}
\begin{tabular}{|m{0.06\textwidth}||m{0.22\textwidth}||m{0.09\textwidth}|}
\hline
\textbf{Notation} & \textbf{Parameter} & \textbf{Value} \\ 
\hhline{|=#=#=|}
$M$ & Number of subchannels & 50\\\hline
$B$ & Number of SBSs & 4, 8\\\hline
$K$ & Number of UEs & 30\\\hline
$N_m$ & Number of antennas for MBS & 64\\\hline
$N_s$ & Number of antennas with SBS & 16\\\hline
$N_u$ & Number of antennas with UE & 2\\\hline
$L$ & Number of NLoS paths & 6\\\hline
$P_0^\mathrm{max}$ &  Maximum transmit power of MBS & 46dBm\\\hline
$P_b^\mathrm{max}$ &  Maximum transmit power of SBS & 30dBm\\\hline
$W$ & Bandwidth per subchannel & 2MHz\\\hline
$N_0$ & Noise variance & -174dBm/Hz\\\hline
$f_c$ & Carrier frequency & 28GHz\\\hline
$\beta_\mathrm{LoS}$ & Path loss exponent of LoS path & 2.1 \\\hline
$\beta_\mathrm{NLoS}$ & Path loss exponent of NLoS path & 3.17\\\hline
$\sigma_\mathrm{LoS}$ & Shadowing standard deviation of LoS path & 3.76dB  \\\hline
$\sigma_\mathrm{NLoS}$ & Shadowing standard deviation of NLoS path & 8.09dB\\\hline
\end{tabular}
\end{center}
\end{table}

\begin{figure}[t]\centering
\subfloat[Case 1 where $B=4$\label{fig:config1}]{\includegraphics[width=45mm]{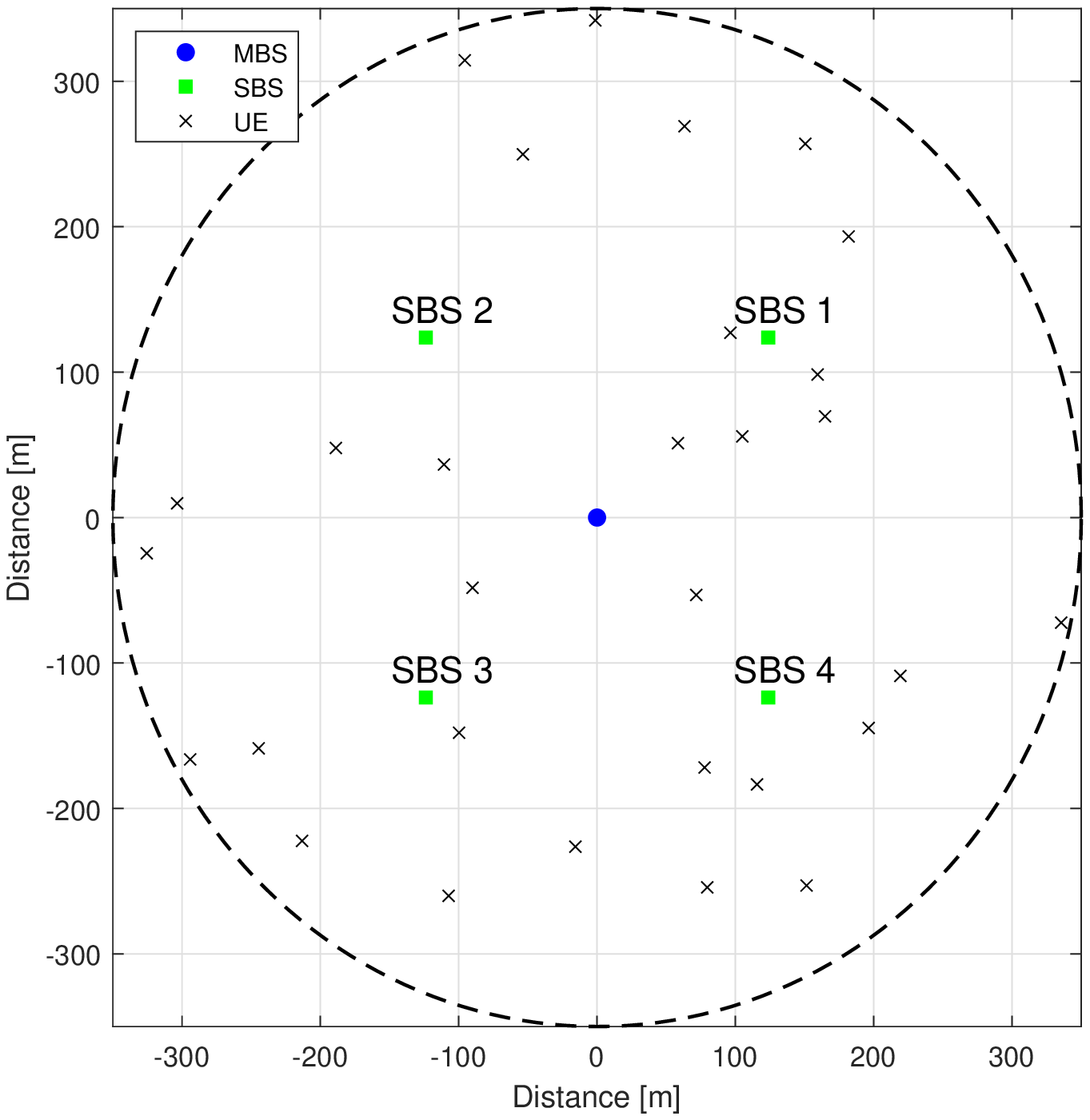}}
\subfloat[Case 2 where $B=8$\label{fig:config2}]{\includegraphics[width=45mm]{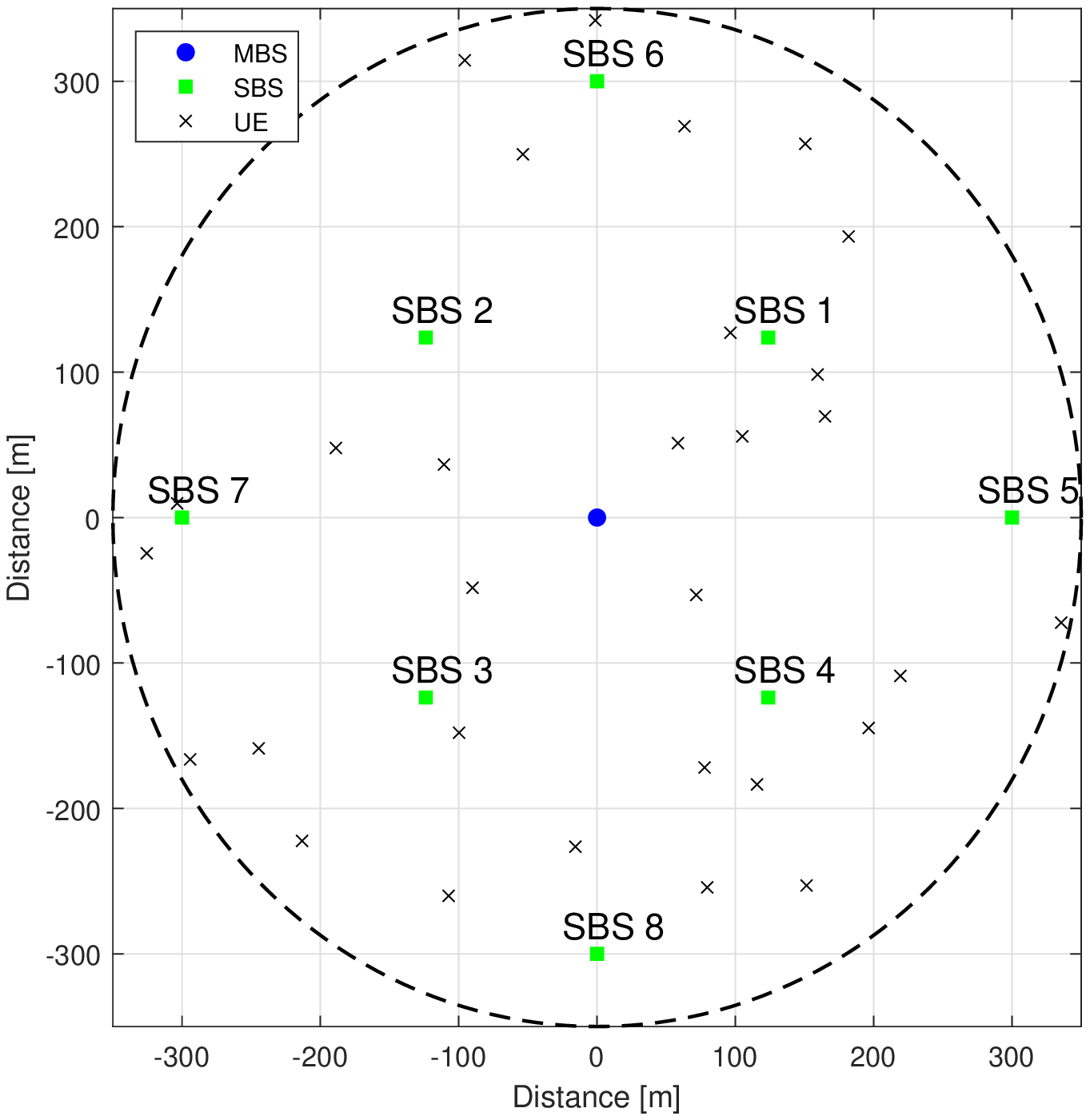}}
 \caption{The deployments of MBS and SBSs for different number of SBSs where dotted line represents the coverage area of MBS}\label{fig:assoc}
\end{figure}
In our simulations, MBS is located at the center with radius 350m and SBSs are located in a fixed position as in \cref{fig:assoc} where 4 SBSs and 8 SBSs are deployed for Case 1 and Case 2, respectively.
We adopt the simulation parameters given in \cref{Simul_table}, unless otherwise specified \cite{PL_model}.
Morever, maximum ratio transmission (MRT) precoding and maximal ratio combining (MRC) are employed for transmitter and receiver, respectively.

\subsection{Snapshot of MH-IAB Network}
\begin{figure}[t]\centering
\subfloat[Case 1\label{fig:conv1}]{\includegraphics[width=80mm,height=60mm]{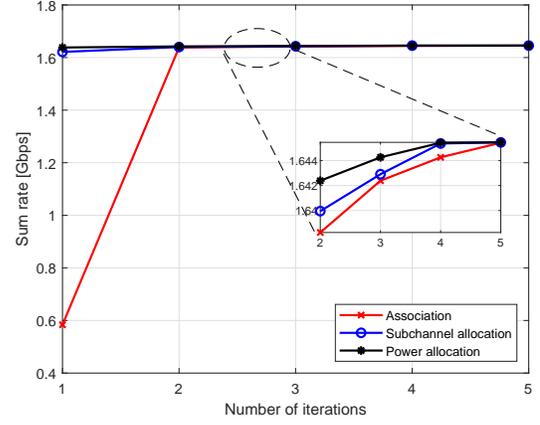}}\\
\subfloat[Case 2\label{fig:conv2}]{\includegraphics[width=80mm,height=60mm]{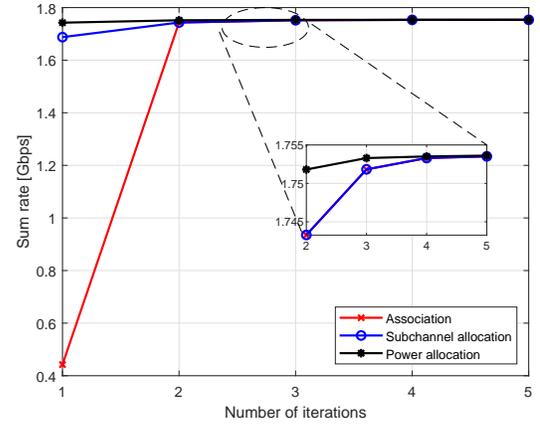}}
 \caption{Convergence behavior for the proposed Algorithm \ref{Algorithm_Main} where $R^\textrm{th}_i=2\textrm{Mbps}$ and $K=30$. }\label{fig:convergence}
\end{figure}
In the case of fixed UE location as in \cref{fig:assoc}, we show the convergence of our proposed Algorithm \ref{Algorithm_Main} in Fig. \ref{fig:convergence} where the optimization results for AO, SA, and PA, i.e., $\{ \mathbf{y}, \mathbf{x}, \mathbf{P} \}$, are getting close in each iteration, as the number of iteration increases, and are converged in the $5$ iterations as seen in \cref{fig:convergence}.
Therefore, we can confirm that the optimization results for $\mathbf{y}$, $\mathbf{x}$, and $\mathbf{P}$ in each step are monotonically increasing, and the joint optimization algorithm converges.


\begin{figure*}[t]\centering
\subfloat[$R^\textrm{th}_i=2$Mbps in Case 1\label{fig:assoc1_1}]{\includegraphics[width=60mm]{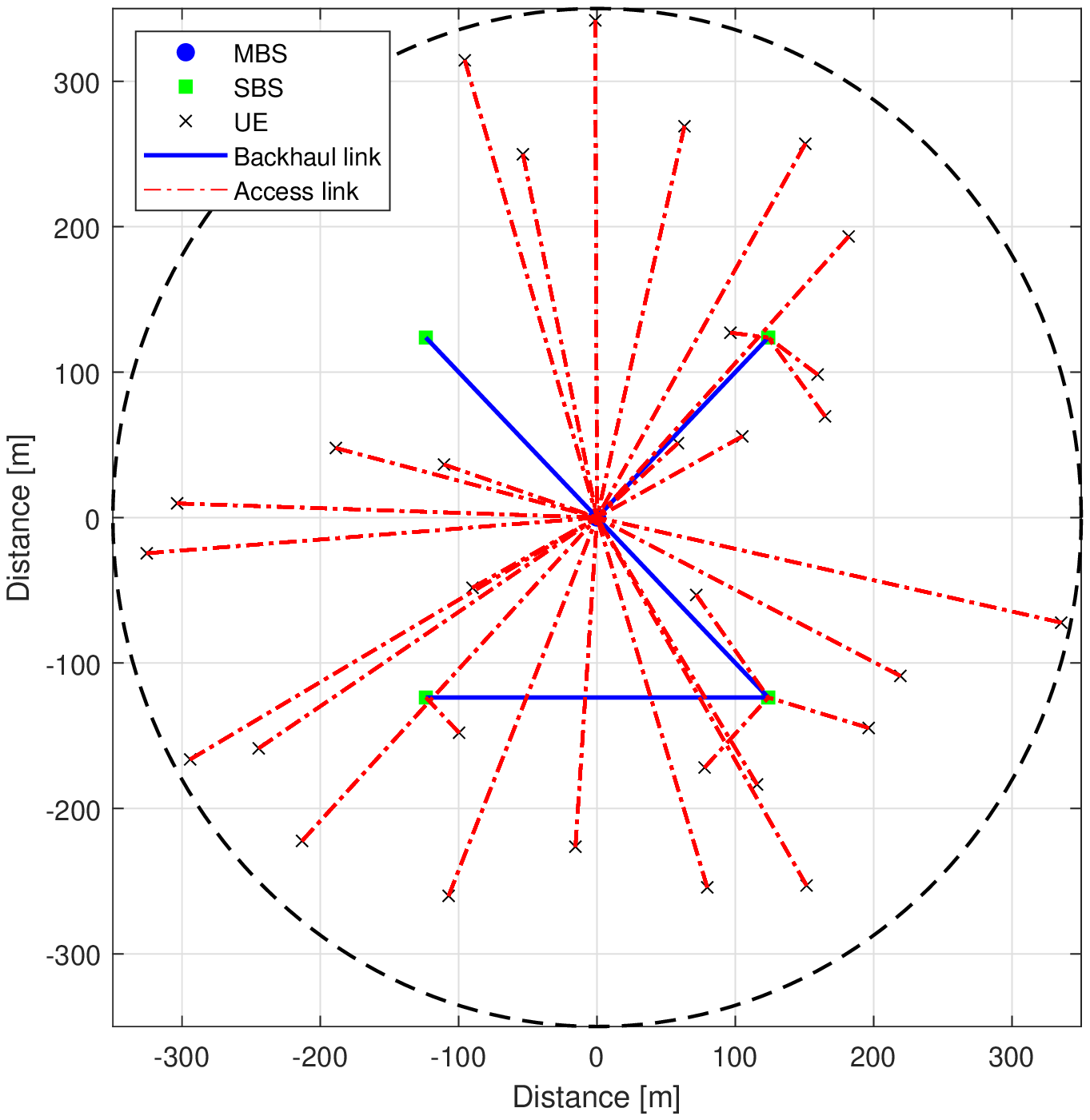}}
\subfloat[$R^\textrm{th}_i=40$Mbps in Case 1\label{fig:assoc1_2}]{\includegraphics[width=60mm]{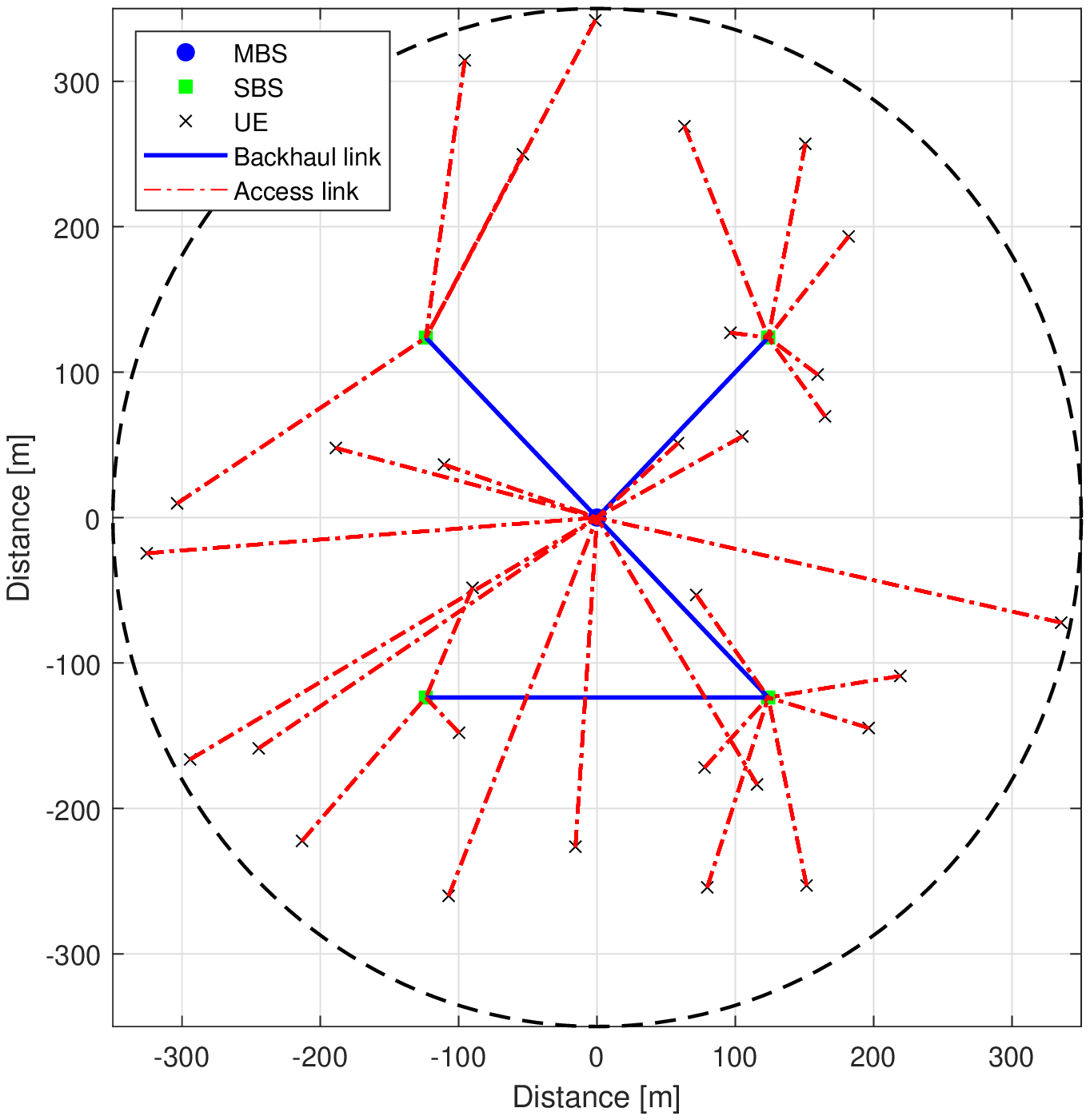}}
\subfloat[$R^\textrm{th}_i=2$Mbps and $P_0^\textrm{max}=30$dBm in Case 1\label{fig:assoc1_3}]{\includegraphics[width=60mm]{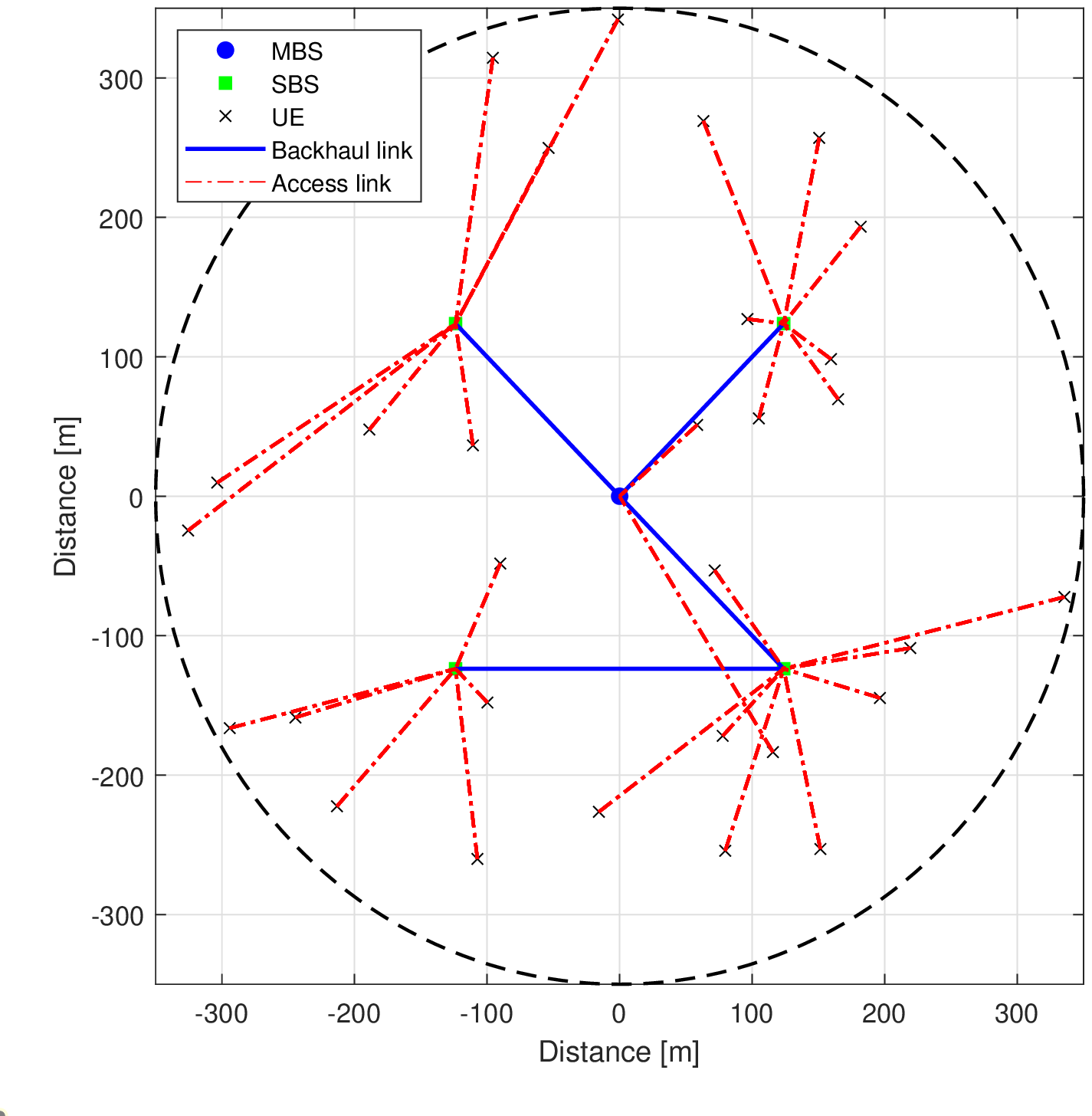}}

\subfloat[$R^\textrm{th}_i=2$Mbps in Case 2\label{fig:assoc2_1}]{\includegraphics[width=60mm]{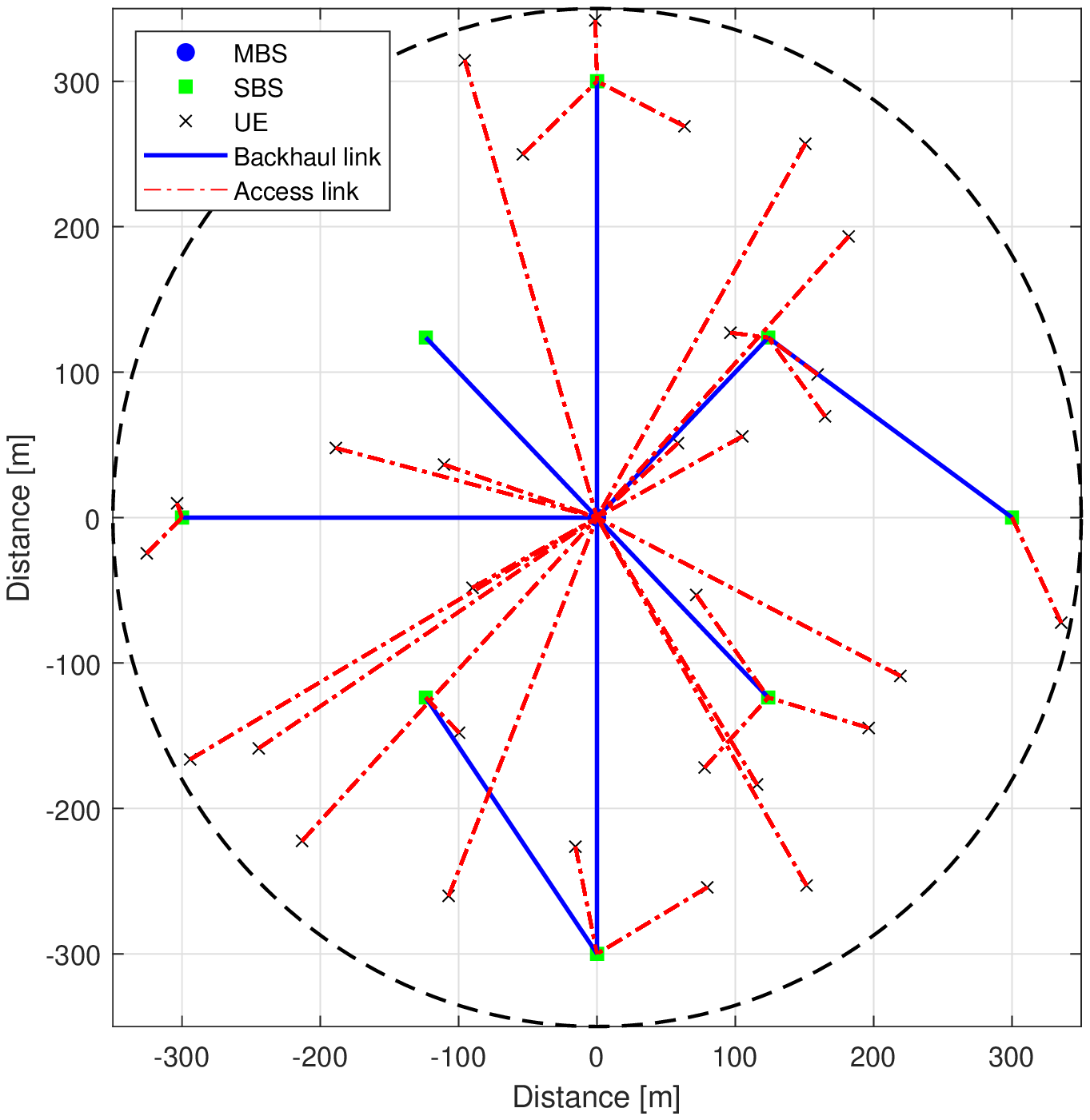}}
\subfloat[$R^\textrm{th}_i=40$Mbps in Case 2\label{fig:assoc2_2}]{\includegraphics[width=60mm]{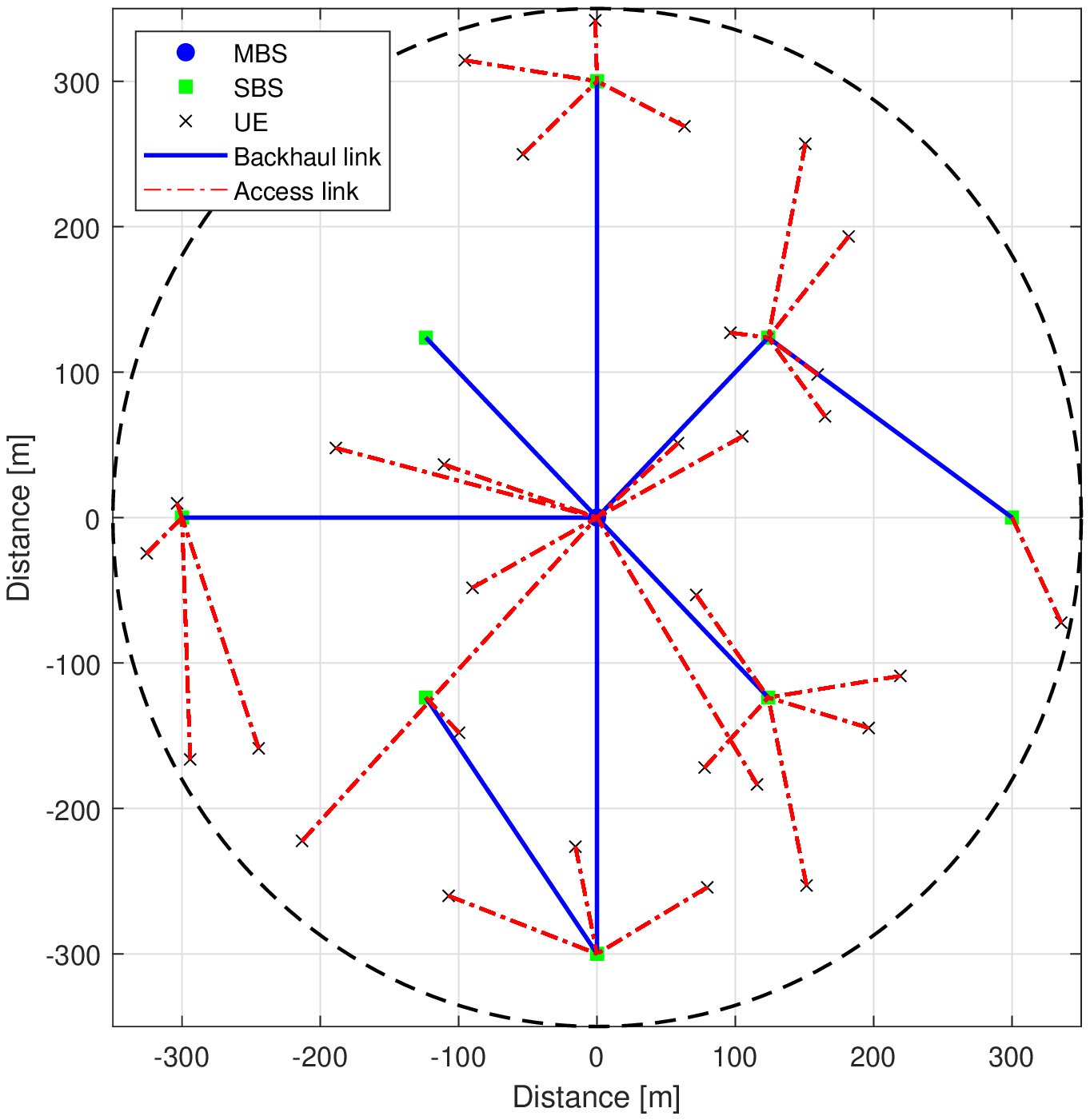}}
\subfloat[{$R^\textrm{th}_i=2$Mbps and $P_0^\textrm{max}=30$dBm in Case 2}\label{fig:assoc2_3}]{\includegraphics[width=60mm]{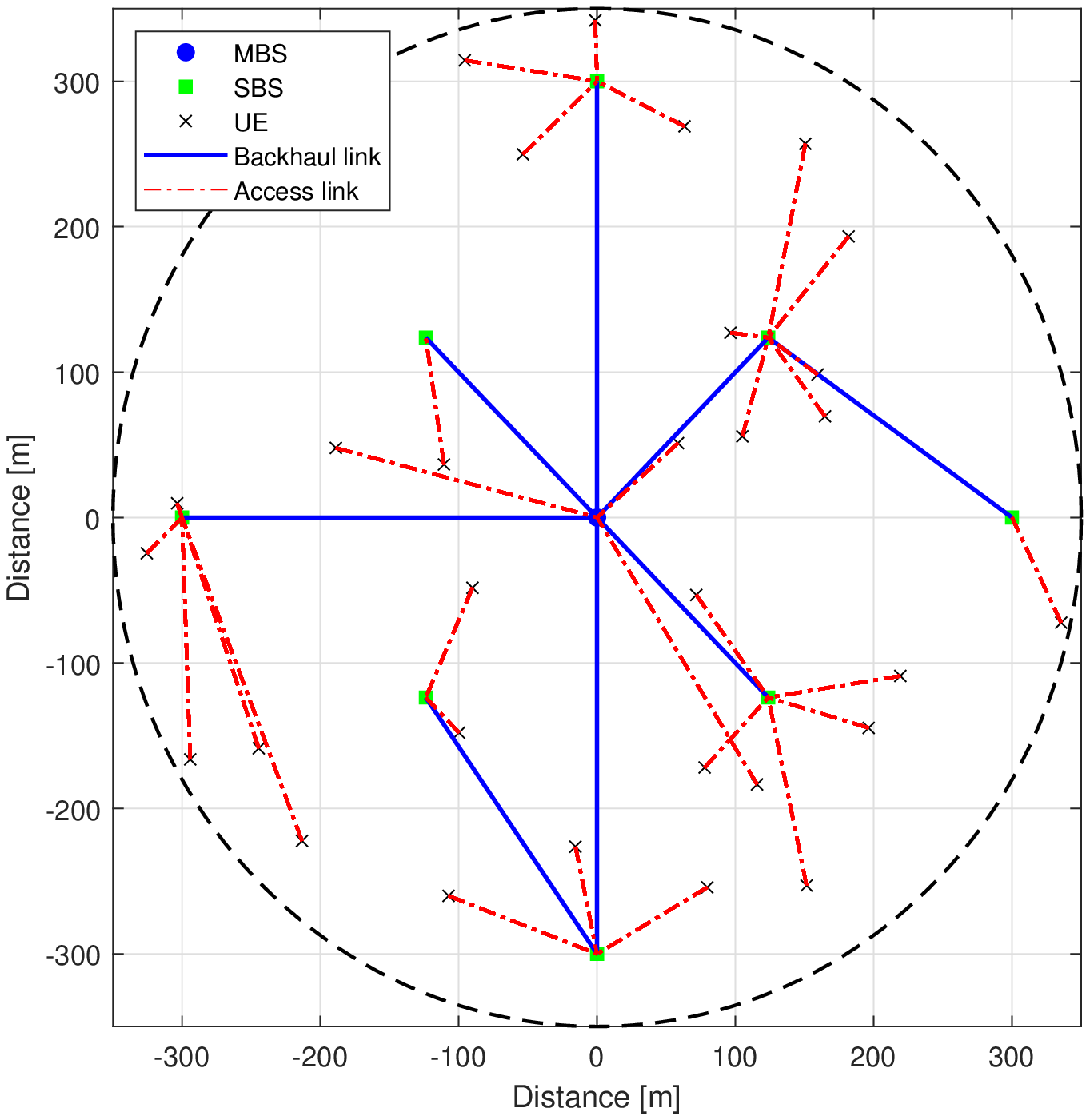}}
 \caption{Examples of AO results obtained by Algorithm \ref{alg1} for various minimum rate requirements and  MBS transmit power where $K=30$ }\label{fig:assoc_result}
\end{figure*}

\cref{fig:assoc_result} illustrates the results of proposed association algorithm under different minimum rate requirements and MBS transmit power for Case 1 and Case 2.
When the data rate requirement is $2$Mbps for an example as in \cref{fig:assoc1_1,fig:assoc2_1}, UEs tend to associate with MBS since it provides a better link quality due to high transmit power and the wired backhaul of MBS is not limited.  
However, in the case that high data rate is required, 
UE become more inclined to access the nearest BSs as long as there is no LoS path for UE. 
Although MBS provides better link quality than SBS, MBS cannot simultaneously support a large number of UEs to meet high data rate requirements using the limited resources.
Thus, it is better to associate UE with lightly loaded SBSs for high data rate requirements to exploit the spatial reuse of limited resources.
Moreover, we demonstrate that most of UEs are associated with the nearest BSs when MBS has the same transmit power with SBS 
as in \cref{fig:assoc2_1,fig:assoc2_3}, while UEs require low data rate. In other words, there is no gain to associate with MBS unless the LoS path is guaranteed.
We also observe that the backhaul links are configured by circumventing NLoS link to enhance the throughput of backhaul. For example, ``SBS 3" connects to ``SBS 8" in Case 2 since the LoS path from `SBS 3' to MBS is blocked.

\begin{table*}[t]
    \centering
    \resizebox{1.45\columnwidth}{!}{\begin{minipage}[t]{1.33\columnwidth}
    \caption{Comparison of LoS probability for backhaul and access links}
    \label{fig:Los}
    \newcolumntype{R}{>{\raggedleft\arraybackslash}X}
\begin{tabularx}{1\linewidth}{l l  l l}
    \toprule[1pt]
     & {Association Scheme} &LoS probability for Case 1 &LoS probability for Case 2 \\   
    \cmidrule(lr){2-2} \cmidrule(lr){3-3} \cmidrule(lr){4-4}
    \multirow{2}{*}{Backhaul Link} & {Single-Hop} & $0.6335$\hspace{3.5pt}  \tikz{
        \fill[fill=color1] (0.0,0) rectangle (1.267,0.2);
        \fill[pattern=north west lines, pattern color=black!30!color1] (0.0,0) rectangle (1.267,0.2); 
    } & $0.4778$\hspace{3.5pt}  \tikz{
        \fill[fill=color1] (0.0,0) rectangle (0.9556,0.2);
        \fill[pattern=north west lines, pattern color=black!30!color1] (0.0,0) rectangle (0.9556,0.2); 
    }\\
    & {\textbf{Multi-Hop}} & $0.8175$\hspace{3.5pt}  \tikz{
        \fill[fill=color2] (0.0,0) rectangle (1.635,0.2);
        \fill[pattern=north west lines, pattern color=black!30!color2] (0.0,0) rectangle (1.635,0.2);
    } & $0.8400$\hspace{3.5pt}  \tikz{
        \fill[fill=color2] (0.0,0) rectangle (1.68,0.2);
        \fill[pattern=north west lines, pattern color=black!30!color2] (0.0,0) rectangle (1.68,0.2);
    }  \\
    & & \ \ \ \ \ \ \ \ \ \tikz{
        \draw[black] (1.1,0) -- (3.1,0);
        \draw[black] (1.1,-2pt) -- (1.1,2pt)node[label=below:\tiny$0$] {};
        \draw[black] (2.1,-2pt) -- (2.1,2pt)node[label=below:\tiny$0.5$] {};
        \draw[black] (3.1,-2pt) -- (3.1,2pt)node[label=below:\tiny$1$] {};
    } & \ \ \ \ \ \ \ \ \ \tikz{
       \draw[black] (1.1,0) -- (3.1,0);
        \draw[black] (1.1,-2pt) -- (1.1,2pt)node[label=below:\tiny$0$] {};
        \draw[black] (2.1,-2pt) -- (2.1,2pt)node[label=below:\tiny$0.5$] {};
        \draw[black] (3.1,-2pt) -- (3.1,2pt)node[label=below:\tiny$1$] {};
    } \vspace{-1mm} \\
    
    \midrule[.8pt]
     & {Association Scheme} & LoS probability for Case 1 &LoS probability for Case 2 \\     
    \cmidrule(lr){2-2} \cmidrule(lr){3-3} \cmidrule(lr){4-4}
    \multirow{3}{*}{Access Link} & {Direct Access} & $0.1446$\hspace{3.5pt}  \tikz{
        \fill[fill=color1] (0.0,0) rectangle (0.2892,0.2);
        \fill[pattern=north west lines, pattern color=black!30!color1] (0.0,0) rectangle (0.2892,0.2);
    } & $0.1446$\hspace{3.5pt}  \tikz{
        \fill[fill=color1] (0.0,0) rectangle (0.2892,0.2);
        \fill[pattern=north west lines, pattern color=black!30!color1] (0.0,0) rectangle (0.2892,0.2);
    } \\
        & {Max SINR} & $0.3246$\hspace{3.5pt}  \tikz{
        \fill[fill=color3] (0.0,0) rectangle (0.6492,0.2);
        \fill[pattern=north west lines, pattern color=black!30!color3] (0.0,0) rectangle (0.6492,0.2);
    } & $0.4740$\hspace{3.5pt}  \tikz{
        \fill[fill=color3] (0.0,0) rectangle (0.9480,0.2);
        \fill[pattern=north west lines, pattern color=black!30!color3] (0.0,0) rectangle (0.9480,0.2);
    } \\
    & {\textbf{Proposed}} & $0.3242$\hspace{3.5pt}  \tikz{
        \fill[fill=color2] (0.0,0) rectangle (0.6484,0.2);
        \fill[pattern=north west lines, pattern color=black!30!color2] (0.0,0) rectangle (0.6484,0.2);
    } & $0.4575$\hspace{3pt}  \tikz{
        \fill[fill=color2] (0.0,0) rectangle (0.9150,0.2);
        \fill[pattern=north west lines, pattern color=black!30!color2] (0.0,0) rectangle (0.9150,0.2);
    } \\
    & & \ \ \ \ \ \ \ \ \ \tikz{
       \draw[black] (1.1,0) -- (3.1,0);
        \draw[black] (1.1,-2pt) -- (1.1,2pt)node[label=below:\tiny$0$] {};
        \draw[black] (2.1,-2pt) -- (2.1,2pt)node[label=below:\tiny$0.5$] {};
        \draw[black] (3.1,-2pt) -- (3.1,2pt)node[label=below:\tiny$1$] {};
    } & \ \ \ \ \ \ \ \ \ \tikz{
       \draw[black] (1.1,0) -- (3.1,0);
        \draw[black] (1.1,-2pt) -- (1.1,2pt)node[label=below:\tiny$0$] {};
        \draw[black] (2.1,-2pt) -- (2.1,2pt)node[label=below:\tiny$0.5$] {};
        \draw[black] (3.1,-2pt) -- (3.1,2pt)node[label=below:\tiny$1$] {};
    } \vspace{-1mm} \\
    \bottomrule[1pt]
\end{tabularx}
    \end{minipage}}
\end{table*}


\subsection{Performance evaluation of IAB network}

Throughout this subsection, our proposed scheme is compared with three alternative schemes as follows.

\begin{enumerate}

\item \textbf{SH with max SINR} is the case where backhaul links are configured by ``Single-Hop" (i.e., SBSs are directly connected to MBS) and access links are configured by ``Max SINR" based association scheme where each UE selects the BS providing the largest SINR link. 
\item \textbf{MH with max SINR} is the case where access links are connected by ``Max SINR" scheme and backhaul links are configured by the proposed MH backhauling.
\item \textbf{SH with prop} implies the case where backhaul links are configured by SH while access links are configured by the proposed association scheme in Algorithm \ref{alg1} given the backhaul links. Note that ``Prop" in the numerical results stands for our proposed joint association scheme for backhaul and access links.
\end{enumerate}
It is noting that the proposed SA and PA algorithms are adopted for all the alternative schemes.
Also, the numerical results are obtained by averaging over the random location of UEs where they are uniformly distributed in a cell.

\begin{figure}[t]\centering
\includegraphics[width=90mm]{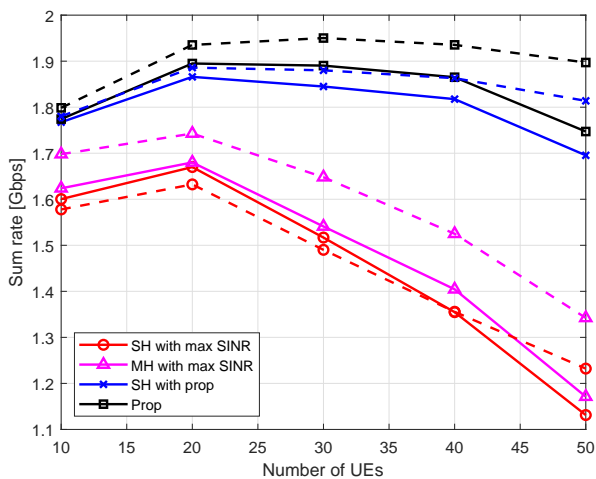}
\caption{Comparison of sum rate for different number of UEs where $R_i^\textrm{th}=2$Mbps. Solid and dotted lines represent Case1 and Case 2, respectively.}\label{fig:sum}
\end{figure}

Table. \ref{fig:Los} represents the LoS probability of backhaul and access links achieved by the baselines and the proposed scheme. 
As seen in Table. \ref{fig:Los}, the LoS probability of backhaul is significantly increased by configuring the MH backhaul links, and particularly it can be further increased when a large number of SBSs are deployed. 
Since each SBS is more likely to obtain the LoS backhaul link under the dense SBS deployment scenario,
the proposed association scheme achieves high LoS probability of backhaul by connecting the LoS path between SBSs. 
For the access link, we compare the proposed scheme with ``Max SINR" and ``Direct Access" that connects all the UEs to MBS.
As in Table. \ref{fig:Los}, the LoS probability of access link can be enhanced using IAB network compared to direct access.
However, the LoS probability of proposed scheme slightly decreases compared with max SINR scheme.
Although some UEs are associated with BS which provides the NLoS link, the proposed scheme achieves better system throughput because access links are configured by taking into account  the backhaul capacity and the load of SBSs as in \cref{fig:sum}.
Therefore, the coverage can be extended using IAB network with the proposed association scheme by providing high LoS probability for both backhaul and access links.

 In \cref{fig:sum}, we compare the MH backauling based proposed scheme with the SH backhauling and max SINR association scheme with respect to the number of UEs.
 We observe that the sum rate increases as the number of UEs increases due to the multi-user diversity in all the schemes. On the other hand, it decreases when a large number of UEs are deployed since it is difficult to satisfy the minimum rate requirements for all the UEs using the limited resources.
 In particular, this phenomenon becomes more severe for max SINR association scheme because it does not consider the limitation of backhaul capacity and the load of SBSs even if the MH backauling is supported.
In contrast, the performance of sum rate using the proposed scheme is significantly boosted compared to max SINR association scheme even if the SH backhauling is used. Since the proposed scheme adaptively configures the access links to resolve the bottleneck of backhaul link, the resources can be efficiently utilized to satisfy the data rate requirements.





\begin{figure}[t]\centering
\includegraphics[width=90mm]{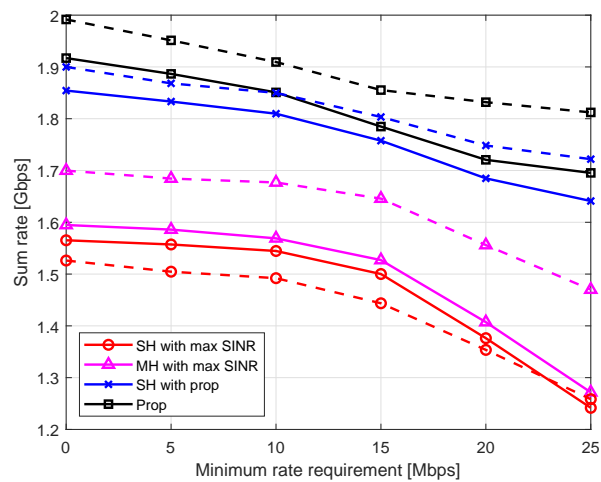}
\caption{Sum rate versus minimum rate requirement where $K=30$. Solid and dotted lines represent Case1 and Case 2, respectively.}\label{fig:min}
\end{figure}

In \cref{fig:min}, we evaluate the sum rate over the minimum rate requirements.
It is shown that the sum rate decreases as the minimum rate requirement increases and the MH backhauling based proposed scheme outperforms the other baselines. 
Thus, the MH backhauling by deploying a large number of SBSs provides a better performance under QoS constraints.
From \cref{fig:sum,fig:min}, we confirm that the proposed scheme achieves better sum rate performance in a dense UE deployment case even though max SINR association scheme achieves higher LoS probability.
Proposed scheme distributedly associates UEs with lightly loaded SBS regardless of backhauling scheme to exploit the spatial reuse of limited spectrum resources.
We here point out that the UE association with the consideration of backhaul condition is one of the important factor to improve the system performance in IAB network.
Additionally, we demonstrate that the sum rate performance can be improved by deploying a large number of SBSs and configuring the MH backhauling regardless of UE association scheme.
However, the dense SBS deployment scenario with ``SH with max SINR" reduces the sum rate performance with the small number of UEs since it provides a low LoS probability of backhaul as in Table. \ref{fig:Los}.



\begin{figure}[t]\centering
\includegraphics[width=90mm]{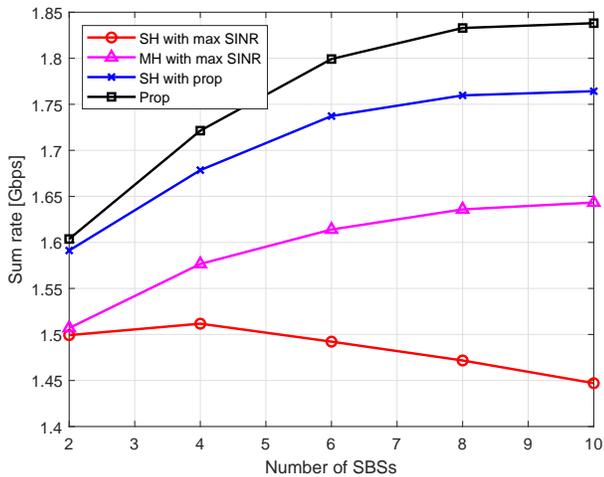}
\caption{Comparison of sum rate performance for different number of SBSs where $R_i^{th}=2$Mbps and $K=30$.}\label{fig:num_SBS}
\end{figure}

\cref{fig:num_SBS} illustrates the sum rate performance with respect to the number of SBSs for different association schemes.
As shown in \cref{fig:num_SBS}, the sum rate increases as the number of SBSs increases by employing the MH backhauling or the proposed association scheme for access links.
On the other hand, ``SH with max SINR" rather deteriorates the system throughput in which more SBSs are deployed, such as ultra-dense network, since it cannot exploit the backhaul conditions.
Here, we notice that the proposed scheme achieves the diversity gain using SBSs. 
Besides, we also highlight that the proposed UE association scheme significantly improves the system throughput
since it associates UEs by considering the backhaul traffic and offloading the traffic from MBS to SBSs.
As a result, we validate that the proposed algorithm supports the MH backhaul connectivity in IAB network, which improves the system throughput by circumventing the blockage and offloading UEs to SBSs when a large number of SBSs are deployed.

\section{Conclusion}\label{conclu}
In this paper, we have investigated the joint association and resource allocation for IAB network in consideration of MH backhauling. 
To maximize the sum rate of IAB network, the joint optimization problem is decomposed into three subproblems. 
First, we used Lagrangian duality approach to address AO for backhaul and access links to configure the MH wireless backhaul links.
Then, by using SCA approach, we tackled the resource allocation including SA and PA.
Simulation results showed that the proposed algorithm provides more spectral-efficient association and it also achieves the diversity gain.
Furthermore, the proposed algorithm for IAB network outperforms the conventional schemes by enabling to use the MH backhauling and to offload UEs from MBS to SBSs.

\section*{Appendix A} 
We prove \emph{Proposition 1} using the abstract Lagrangian duality.
The primal solution of \eqref{P3.2} is given as
\begin{align}
    p^*=\underset{\mathbf{x}}{\max}~\underset{\mu}{\min} ~L(\mathbf{x},\mu)
\end{align}
Also, the solution of dual problem can be written as
\begin{align}\label{dual_solution}
    d^*=\underset{\mu}{\min}~\underset{\mathbf{x}}{\max}~ L(\mathbf{x},\mu)=\underset{\mu}{\min}~ \theta(\mu)
\end{align}
where $\theta(\mu)=\underset{\mathbf{x}}{\max}~ L(\mathbf{x},\mu)$.
According to the weak duality, $d^*\geq p^*$ always holds.
Note that we have $\sum_{b\in \mathcal{B}_0}\sum_{i\in \mathcal{I}}\sum_{m \in \mathcal{M}}(x_{b,i,m}-x_{b,i,m}^2)\geq 0$ for $\mathbf{x}$ such that $\bar{C}_1, \bar{C}_2, C_3, C_4$, and $C_{11a}$.
Thus, two cases can be considered 

\emph{Case 1}: Let us consider $\sum\limits_{b\in \mathcal{B}_0}\sum\limits_{i\in \mathcal{I}}\sum\limits_{m \in \mathcal{M}}(x_{b,i,m}-x_{b,i,m}^2)=0$.
In this case, $L(\mathbf{x},\mu)=\underset{\mu}{\min}~ L(\mathbf{x},\mu)$ holds.
Also, the following inequality holds
\begin{align}\label{dual}
    d^*=\underset{\mu}{\min}~\underset{\mathbf{x}}{\max}~ L(\mathbf{x},\mu)\leq \underset{\mathbf{x}}{\max}~ L(\mathbf{x},\mu)
\end{align}
Substituting $L(\mathbf{x},\mu)=\underset{\mu}{\min}~ L(\mathbf{x},\mu)$ into the right side of \eqref{dual} yields
\begin{align}\label{dual2}
    \underset{\mu}{\min}~\underset{\mathbf{x}}{\max}~ L(\mathbf{x},\mu)\leq \underset{\mathbf{x}}{\max}~ \underset{\mu}{\min}~ L(\mathbf{x},\mu)
\end{align}
By comparing the weak duality condition with \eqref{dual2}, the strong duality holds as follows.
\begin{align}
    d^*=\underset{\mu}{\min}~\underset{\mathbf{x}}{\max}~ L(\mathbf{x},\mu)= \underset{\mathbf{x}}{\max}~ \underset{\mu}{\min}~ L(\mathbf{x},\mu)=p^*
\end{align}
Note that $\theta(\mu)$ is monotonically decreasing function with respect to $\mu$.
Since $d^*=\underset{\mu}{\min}~\theta(\mu)$ as in \eqref{dual_solution}, we conclude that $\theta(\mu)=d^*$ for $\mu\geq \mu^*$ where $\mu^*=\underset{\mu}{\arg}~ \min~\theta(\mu)$.

\emph{Case 2}: For the case of $\sum\limits_{b\in \mathcal{B}_0}\sum\limits_{i\in \mathcal{I}}\sum\limits_{m \in \mathcal{M}}(x_{b,i,m}-x_{b,i,m}^2)>0$, $\theta(\mu^*)$ tends to $-\infty$ due to monotonically decreasing function of $\theta(\mu)$. However, it contradicts the weak duality since the lower bound of $\theta(\mu)$ is $p^*$ which is always greater than zero.
Therefore, $\sum\limits_{b\in \mathcal{B}_0}\sum\limits_{i\in \mathcal{I}}\sum\limits_{m \in \mathcal{M}}(x_{b,i,m}-x_{b,i,m}^2)=0$ holds at the optimal point.

\bibliographystyle{IEEEtran}
\bibliography{IEEEabrv,database}

\end{document}